\documentclass[acmsmall,table,xcdraw]{acmart}

\usepackage{multirow}
\usepackage{caption}
\usepackage{graphicx}
\usepackage{enumitem} 
\usepackage{soul} 
\usepackage{xspace} 
\usepackage{subcaption} 
\usepackage{multirow} 
\usepackage{siunitx} 
\usepackage[many]{tcolorbox} 
\usepackage{xurl} 
\usepackage{listings}
\AtBeginDocument{\DeclareCaptionSubType{lstlisting}}
\usepackage{caption,subcaption}

\AtBeginDocument{%
  \providecommand\BibTeX{{%
    \normalfont B\kern-0.5em{\scshape i\kern-0.25em b}\kern-0.8em\TeX}}}

\definecolor{codegreen}{rgb}{0,0.6,0}
\definecolor{codegray}{rgb}{0.5,0.5,0.5}
\definecolor{codepurple}{rgb}{0.58,0,0.82}
\definecolor{backcolour}{rgb}{0.95,0.95,0.92}

\lstdefinestyle{mystyle}{
    backgroundcolor=\color{backcolour},   
    commentstyle=\color{codegreen},
    keywordstyle=\color{magenta},
    numberstyle=\tiny\color{codegray},
    stringstyle=\color{codepurple},
    basicstyle=\ttfamily\footnotesize,
    breakatwhitespace=false,         
    breaklines=true,                 
    captionpos=b,                    
    keepspaces=true,                 
    numbers=left,                    
    numbersep=5pt,                  
    showspaces=false,                
    showstringspaces=false,
    showtabs=false,                  
    tabsize=2
}

\acmJournal{TOSEM}

\newcommand{\Lionel}[1]{{\color{red}{[Lionel: #1]}}}
\newtcolorbox{mybox}[2][]{
top=0.15in,left=4pt,right=4pt,bottom=4pt,
fonttitle=\bfseries,
colbacktitle=gray,
colback=gray!5,
colframe=gray!40!black,
enhanced,
attach boxed title to top left={xshift=1.5em,yshift=-\tcboxedtitleheight/2},
boxed title style={size=small},
drop shadow={black!50!white},
title=#2,#1}

\newcommand{\countobservations}{
    \def \countobservations{1}
}
\newcounter{observation}
\countobservations

\newcommand{\countimplications}{
    \def \countimplications{1}
}
\newcounter{implication}
\newcommand{\subsubsubsection}[1]{\paragraph{#1}\mbox{}\\}

\countimplications

\newcommand\RQOne{\textbf{How is the usage of deep learning libraries evolving over time?}}
\newcommand\RQTwo{\textbf{Do developers use multiple DL libraries in software projects?}}
\newcommand\RQThree{\textbf{What are the most prevalent combinations of DL libraries across ML pipeline?}}

\makeatletter
\def\l@paragraph{\@tocline{4}{0pt}{1pc}{7pc}{}}
\def\l@subparagraph{\@tocline{5}{0pt}{1pc}{7pc}{}}
\@xp\gdef\csname r@tocindent4\endcsname{0pt}
\makeatother
\stepcounter{secnumdepth}
\stepcounter{tocdepth}

\newlength\MAX  
\newcommand*\Chart[1]{#1~\rlap{\textcolor{black!20}{\rule{\MAX}{2ex}}}\rule{#1\MAX}{2ex}}

\graphicspath{{figures/}}

\begin{document}

\title{An Empirical Study of Library Usage and Dependency in Deep Learning Frameworks}

\author{Mohamed Raed El aoun }
\email{mohamed-raed.el-aoun@polymtl.ca}
\affiliation{%
  \institution{SWAT Lab., Polytechnique Montreal}
  \city{Montreal}
  \country{Canada}
}

\author{Lionel Nganyewou Tidjon}
\email{lionel.tidjon@polymtl.ca}
\affiliation{%
  \institution{SWAT Lab., Polytechnique Montreal}
  \city{Montreal}
  \country{Canada}
}

\author{Ben Rombaut }
\email{mohamed-raed.el-aoun@polymtl.ca}
\affiliation{%
  \institution{SAIL Lab., Queen's University}
  \city{Kingston}
  \country{Canada}
}

\author{Foutse Khomh}
\email{foutse.khomh@polymtl.ca}
\affiliation{%
  \institution{SWAT Lab., Polytechnique Montreal}
  \city{Montreal}
  \country{Canada}
}

\author{Ahmed E. Hassan}
\email{ahmed@cs.queensu.ca>}
\affiliation{%
  \institution{SAIL Lab., Queen's University}
  \city{Kingston}
  \country{Canada}
}

\renewcommand{\shortauthors}{M. Raed El Aoun et al.}

\begin{abstract}
Recent advances in Deep Learning (DL) have led to the release of several DL software libraries such as PyTorch, Caffe, and TensorFlow, in order to assist Machine Learning (ML) practitioners in developing and deploying state-of-the-art Deep Neural Networks (DNN), but they are not able to properly cope with limitations in the DL libraries such as testing or data processing. In this paper, we present a qualitative and quantitative analysis of 
the most frequent DL libraries combination, the distribution of  DL library dependencies across the ML workflow, and formulate a set of recommendations to (i) hardware builders for more optimized accelerators and (ii) library builder for more refined future releases. Our study is based on 1,484 open-source DL projects with 46,110 contributors selected based on their reputation. First, we found an increasing trend in the usage of deep learning libraries. Second, we highlight several usage patterns of deep learning libraries. In addition, we identify dependencies between DL libraries and the most frequent combination where we discover that PyTorch and scikit-learn and, Keras and Tensorflow are the most frequent combination in 18\% and 14\% of the projects. The developer uses two or three DL libraries in the same projects and tends to use different multiple DL libraries in both the same function and the same files. The developer shows patterns in using various deep learning libraries and prefers simple functions with fewer arguments and straightforward goals. Finally, we present the implications of our findings for researchers, library maintainers, and hardware vendors.

\end{abstract}

\ccsdesc[500]{General and reference~Empirical studies}
\ccsdesc[500]{Software development trend}
\ccsdesc[500]{Computer systems organization~Deep learning}

\keywords{Deep learning libraries, empirical studies, dependency model}

\maketitle

\section{Introduction}
\label{sec:introduction}
Deep learning (DL) is a subset of machine learning (ML) which in turn is a subset of artificial intelligence (AI) which comes with idea to imitate human capabilities by a computer. In deep learning, engineers gather training data and simply feed it into the DL algorithm (artificial neuron network) that will find out an approximation of the function that representing a definition of the training data. Formally, the engineer first expresses their needs, then designs an algorithm to solve a problem, and finally puts together a complex system made from smaller components. Instead of coding the component, in DL everything is about data where engineers replace compilation by training a model with a training dataset. In addition to approximated functions by deep learning (DL) algorithms, engineers still rely on traditional coding to implement DL algorithms. Deep learning has shown big growth in the past years and gained more popularity. Program using deep learning algorithms has shed light on many new inventions and domain applications such as self-driving cars~\cite{10.1145/3180155.3180220}, medical software analysis~\cite{DBLP:journals/corr/LitjensKBSCGLGS17}, drug discovery~\cite{CHEN20181241}, fraud detection~\cite{8632885} and language translation~\cite{8632885}. Depp learning (DL) libraries such as \texttt{Tensorflow}, \texttt{PyTorch}, \texttt{Keras}, \texttt{scikit-Learn} have a big contribution for the growth of deep learning (DL). Pioneers like Google, Facebook, and Microsoft are relentlessly releasing their DL products, and DL frameworks, and continuously investing in open-source ML libraries. The easy access to the DL libraries makes it easy for researchers to release DL algorithms with minimal effort. Even with the help of DL libraries, today, with the help of DL libraries, developers can make applications learn from scratch developers can easily deep learn applications from scratch or even transform traditional software into ML-system if they have access to training data. 

With the growing interest in deep learning (DL), we observe a rapid evolution of this domain. Researchers are actively designing new DL algorithms and relentlessly trying to optimize the existing DL algorithms. Therefore, deep learning libraries are evolving rapidly releasing new versions and new DL libraries are getting released. Moreover, DL libraries require many computations, which calls for optimized hard wares such as GPUs and TPUs to efficiently train a model. Because of this, the unique characteristics of usage of DL libraries need careful consideration from researchers. 


Researchers have extensively studied traditional libraries~\cite{6676899},~\cite{10.1145/2393596.2393661}.~\cite{,10.1145/3133956.3134059},~\cite{6747226}.~\cite{10.5555/1133105.1133107},~\cite{10.1145/3236024.3275535},~\cite{DBLP:journals/corr/abs-1710-04936},~\cite{10.1007/s10664-014-9325-9}, but there are few studies that answer fundamental questions about the usage of deep learning. Islam and al~\cite{10.1145/3338906.3338955} studied the GitHub issue reports to identify bug characteristics in ML applications. Han and al~\cite{ef3d587226ff4ddc9cf57dc9608bbc7c} studied the discussion topics in StackOverflow posts while Dilhara and al~\cite{10.1145/3453478} study the ML library evolution and updates. None of the cited works answer fundamental questions such as how are the most used DL libraries usage evolves over time? Does practitioner use multiple DL libraries and Why? What are the opportunities to assist library maintainers and Hardware builders?. This lack of information impacts (1) researchers who are not aware of the gap (challenges that developers are facing) and miss opportunities to improve the state of the art of deep learning. (2) library builders are not aware of the challenges that their users are facing and if their customers are properly exploiting the library. (3) Hardware builders are not aware of what to properly optimize on accelerators. (4) Developers are left in the dark about the good practices of the DL libraries.

To fill this gap, we present a qualitative and quantitative empirical study to understand how deep learning practitioners use deep learning libraries. Using static analysis, we study 1,484 open source deep learning projects written in python with 46,110 contributors. Using this rich data, we answer the following questions:
\begin{description}
\item [\textbf{RQ1:}] \RQOne{} In this research question, we analyze the source code of the projects across all it is versions and study the popularity of DL libraries over time. The aim being to identify the most popular DL library and the evolution of it's usage over time. We find that between 2015 and 2019 DL library usage has grown from 2\% to 80\%. \texttt{Tensorflow}, \texttt{PyTorch} and \texttt{Keras} libraries are the most popular DL library among practitioners. However, \texttt{Tensorflow} is the most popular deep learning library, but \texttt{PyTorch} starts to gain popularity since 2019 while \texttt{Tensorflow} popularity is decreasing. In addition, most of the project that uses DL libraries are popular according to their number of stars and forks. Also, most of the projects that use AutoML tools are relatively old based on age.

\item [\textbf{RQ2:}] \RQTwo{} In this research question, we perform a code static analysis of the project that uses at least two deep learning libraries and highlight different dependency models between DL libraries to prove the dependency between DL libraries and identify the different used combinations. We observed that 45.85\% of the projects use at least two DL libraries to implement various stages of the ML workflow. In particular, practitioners tend to use a combination of 2 and 3 DL libraries in the same parent function or in the same file during the life cycle of the project.

\item [\textbf{RQ3:}] \RQThree{} In this research question, we extract the different code patterns that use multiple deep learning libraries across ML workflow (i.e. Model requirements, data collection, data cleaning, feature engineering, model training, model evaluation, model deployment and model monitoring) and we analyze the function calls to understand the reasons behind the dependency. We observe that practitioners prefer to use function with one job and tend to use function with less arguments which result into using different DL libraries. In addition, the most frequent dependency is between \texttt{Keras}$\Leftrightarrow$\texttt{Tensorflow} and \texttt{Scikit$-$learn} $\Leftrightarrow$ \texttt{PyTorch} in the \texttt{data collection} stage, while in the Feature engineering stage the dependency between \texttt{PyTorch} $\Leftrightarrow$ \texttt{TensorFlow} and \texttt{Scikit$-$learn} $\Leftrightarrow$ \texttt{PyTorch} is the most recurrent.
\end{description}

Based on our findings, we highlight several recommendations: new DL library opportunities to better assist the deep learning community, opportunities to improve existing deep learning libraries according to
the requirements of Practitioners, common practices of deep learning practitioners in the usage of deep learning libraries that help hardware builders to better optimize the accelerators.

The rest of the paper is organized as follows. In Section~\ref{sec:Background}, we discuss the background about deep learning libraries and machine learning workflow. Section~\ref{sec:methodology} describes the design of our study.
Section~\ref{sec:results}, we present our results. Section~\ref{sec:discussion} discusses the implication of our results. Section~\ref{sec:related-work} reviews the related work.
Section~\ref{sec:threats} discusses threats to the validity of our findings. Finally, Section~\ref{sec:conclusion} concludes the paper.
\section{Background}
\label{sec:Background}
In the machine learning pipeline, deep learning libraries are used in one or more stages (i.e., feature engineering, model evaluation). In this section, we discuss the nine stages of the machine learning pipeline and the deep learning libraries. In this study, we refer to deep learning developers, a researcher, data scientists, and deep learning engineers as "DL practitioners".
\subsection{Deep Learning Libraries}

Deep learning is a subset of machine learning, with the main goal being to simulate the behavior of the human brain with algorithms called \texttt{artificial neural network} with three or more layers. Deep learning proposes several learning algorithms such as supervised, unsupervised, and reinforcement learning. To have fast access to these algorithms, the deep learning community has provided DL practitioners with numerous open-source DL libraries such as Scikit-Learn~\cite{scikit-learn}, TensorFlow~\cite{tensorflow2015-whitepaper}, Keras~\cite{chollet2015}, PyTorch~\cite{NEURIPS2019_9015}, Theano~\cite{2016arXiv160502688short} and Caffe~\cite{jia2014caffe}. These libraries provide practitioners with ready-made algorithms or the necessary tools to build algorithms that give hand to easily apply DL-based solutions in the real world. DL libraries come with unique challenges because of the data dependencies, already trained models, and the need for performant hardware. Moreover, DL practitioners often use different DL libraries for different ML workflow stages. For example, DL practitioners use libraries with optimized numerical computation like \texttt{Numpy} and  \texttt{Pandas} for data manipulation, while using \texttt{Matplotlib} for visualisation. 


Because the Deep learning community provides variety DL libraries with many similarities, and DL libraries bring unique challenges, in this study, we look to uncover the dependency between the DL libraries and understand the reasons behind it. 

\subsection{Machine Learning Workflow}

\begin{figure}[!t]
\centering
\captionsetup{justification=centering}
    \centering
    \includegraphics[width=0.8\textwidth]{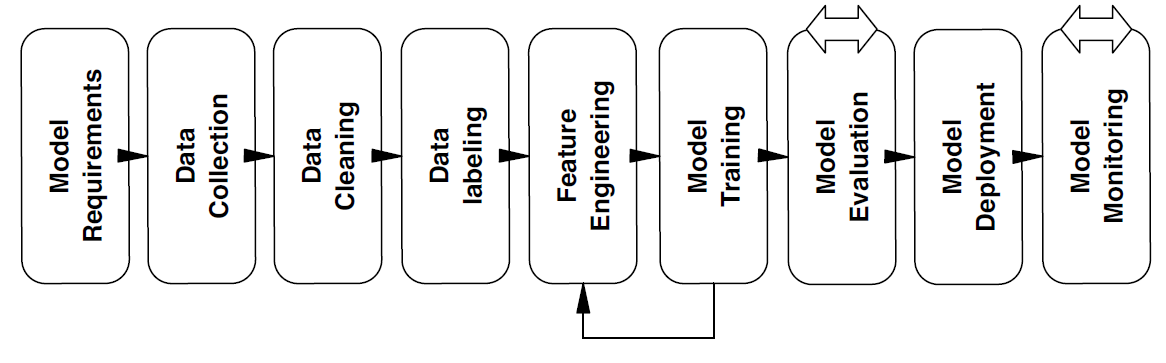}
    \caption{The stages of the ML workflow~\cite{8804457}}
    \label{fig:ML_workflow}
\vspace{-3mm}
\end{figure}

Amershi and al.~\cite{8804457} describes the machine learning pipeline in nine stages for building and deploying ML software systems. ML pipelines are adopted by pioneers such as Google~\cite{google}, Microsoft~\cite{8804457}, and IBM~\cite{IBM}. In this section, we describe the different ML pipeline stages following the reference~\cite{8804457}. Amershi and al group the ML pipeline stages into data and model-oriented categories. \textbf{\texttt{Data-oriented:}} stages are as follow: (1) \texttt{Model requirements} (identifying the most suitable model for the given problem and which feature is appropriate for the model) , (2) \texttt{Data collection} (collecting integrated data sets or generate their own), (3) \texttt{Data cleaning} (removing inaccurate and noisy data records), (4) \texttt{Data labeling} (assign ground truth label to each data record), and (5) \texttt{Feature engineering}(identify and select impactful and informative feature for the appropriate model). \textbf{\texttt{Model-oriented:}} (1) \texttt{Model training} (training and tuning the model using the selected, cleaned, and labeled features), (2) \texttt{Model evaluation} (Evaluate the output of the trained model on a test and safeguard data sets), (3) \texttt{Model deployment} (Deploy the metadata information of the model into the production environment) and (4) \texttt{Model monitoring} (Observe the model behavior in the production environment)
\section{Research-methodology}
\label{sec:methodology}
In this study we identify the dependency between DL libraries and understand reasons behind it. We look to help improve DL libraries future releases and better optimize the effort for hardware builders. We used quantitative and qualitative analysis to answer \texttt{three} research questions. First, we analyse the corpus of 3,134 projects that were identified as DL projects software based on their popularity and activity. Also, we conduct a qualitative study by surveying $n$ developers who introduced multiple DL library in a project from our set. 

The qualitative analysis highlights how developers use DL libraries and how the dependency evolves over the project lifetime, mean while, the qualitative study helps us understand the reasons, challenges and cost with using multiple DL libraries. Next, the quantitative study motivates the qualitative study, in fact the quantitative analysis result has shed light to various research questions that we can answer in depth.
\subsection{Subject Systems}
\label{subsec:Systems}
\subsubsection{DL python projects} Our project set contains 18,122 popular Python deep learning projects hosted in GitHub. We follow Yu et al~\cite{DBLP:journals/corr/abs-1805-01965} for the project selection. Following prior work, we extracted the non-forked and non-archived python projects with a popularity based on the stargazer count~\cite{7816479}, \cite{BORGES2018112} are larger than 50 as it stands at \texttt{September 31,2021}. Inactive projects were discarded. In fact, every project project that does not have an activity prior to our study data is considered inactive. Moreover referring to the advice of Kalliamvakou et al~\cite{Kalliamvakou}, from the active projects we consider only the repositories with more than 3 contributors. Finally we ended up with 3,143 projects. This sample size represents a confidence level of 99\% with 2.3 margin error of the initial population size.

The final dataset is diverse regarding the application domain, size and testing practices. Also, the projects cover a different domains such as framework, robotics, natural language processing and image processing. Having a diverse dataset helps our study to be more representative.

\subsubsection{DL python categories}
\label{subsubsec:projectcategories}
In order to further discover the dependencies between the DL libraries, we have defined four categories to highlight the degree of the dependencies. Each project is going to be assigned a category based on his dominant or most used DL library during his life cycle. To assign the project to his relative category, for each project $p \in{P}$, we look into the static source code of the final version $v_f \in{V}$, where $P$ is a set of studied projects and $V$ a set of versions of the studied projects. If the number of lines of code of $p$ in $v_f$ is dominated by library name $l$ $\in{L}$ then the project is considered as the category \textbf{$l$-Based Project}, where $L$ is the set of all libraries used the studied projects.
We have obtained four categories representing our set of projects \textit{Tensorflow Based Projects, PyTorch Based Projects, Keras Based Projects and, Scikit-Learn }
\subsection{Static Source Code Analysis}
\label{subsec:codeanalysis}
In order to detect the clients of DL libraries and their usage, we analyze all the versions of each project in our corpus.

\subsubsection{Identifying DL applications}
\label{subsubsec:DLapplication}
To study the dependency of DL library usage, it is important to identify the project that uses deep learning library(s). We examine the major releases of the DL libraries and retrieved all the root names along with the most popular alias as $R$. To decide if a project uses a DL library we check (1) for each project $p \in{P}$  we check if it contains an import statement that points to a root name $r \in{R}$. (2) The project must invoke at least one call from the DL library(s). We detect 3,143 projects hosted on GitHub that use at least one DL library $l \in{L}$.  

\subsubsection{Analyse the API Usage}
\label{subsubsec:APIusage}
We perform static code analysis on the 3143 projects to understand how developers use the APIs provided by the DL libraries by identifying the method invocation and the referenced objects that are called from the DL library. We use the \texttt{Abstract Syntax Tree (AST)} module in python to parse the corpus of the project as follow : 
\begin{itemize}
    \item For each version $v \in{V}$ of $p \in{P}$, we extracted the import statement $I$ of every DL library $l \in{L}$ at every file $f$ to define a tuple $t_{import} =\langle p, v, f, I, l \rangle$.
    \item For each version $v \in{V}$ of $p \in{P}$, we extracted the call statement $c$ of every DL library $l \in{L}$ at every file $f$ to define a tuple $t_{call} =\langle p, v, f, c, l \rangle$.
\end{itemize}

\subsection{Dependency model}
\label{subsubsec:DModel}
We define the dependency model as a tuple $t_{dependency}$ of a list of DL libraries $l$ used by a project $p$ at any time of it is life cycle. To identify the DL library dependency, we follow three strategies. First, we identify the project that uses multiple libraries in separate files. Second, we look for projects that use different libraries in the same files but with separate function. Finally, we identify project that uses multiple DL libraries in the same function. This hierarchy can give us an insight on the degree of the dependency between the DL libraries. 
\subsubsection{Identifying Dependent DL libraries at project level}
\label{subsubsec:DLproject}
In this level, for each project $p \in{P}$ at a version $v \in{V}$ if a project used two or more DL libraries $l \in{L}$ in separate files, the project is added to this level. Therefore, our dependency model is a tuple $t_{dependency}=\langle p, v, l, L \rangle$. 

\subsubsection{Identifying Dependent DL libraries at file level}
\label{subsubsec:DLfile}
In this level, for each project $p \in{P}$ at a version $v \in{V}$ if a project used two or more DL libraries $l \in{L}$ in the same file $f$, the project is added to this level. Therefore, our dependency model is a tuple $t_{dependency}=\langle p, v, l, f, L \rangle$. 

To assign each project to his correspondent dependency level, We analyse the source code of each project $p \in{P}$ at each version v $\in{V}$. 
\begin{itemize}
    \item \texttt{Project level dependency}: When a project has multiple DL libraries in different files. 

    \item \texttt{File level dependency} When project has multiple DL libraries in the same file but in different functions.
    
    \item \texttt{Function level dependency} When project has multiple DL libraries in the same function.
\end{itemize}

\subsubsection{Identifying Dependent DL libraries at function level}
\label{subsubsec:DLfunction}
In this level, for each project $p$ $\in{P}$ at a version $v \in{V}$ if a project  used two or more DL libraries $l \in{L}$ in the same file $f$  and same function $F$, the project is added to this level. Therefore our dependency model is a tuple $t_{dependency}=\langle p, v, l, f, F, L\rangle$.

\subsubsection{Identifying Dependent DL libraries across ML development workflow}
\subsection{Manual analysis}
\label{subsec:manualanalysis}
Following previous work~\cite{article.1515123}, two authors (i.e, coders) performed the manual analysis independently. Each coder reads the documentation of the method and assigns two labels. In order to have a consistent labeling strategy, after labeling 25\% of the data, we had a meeting to discuss the results in the first round and reached an agreed-upon the labels. During this phase, the third author of the paper is involved in the discussion. After reaching a 80\% inter-coder agreement, the two coders resumed assigning labels in the rest of the dataset. (recommended inter-coder agreement is between 70\% to more than 90\% in the literature~\cite{doi:10.1177/0049124113500475}). After the second round, we had a meeting to discuss the labeling results, validate labels and verify the consistency of our labels. A third author is also involved in the discussion. Based on the second-round discussion, each coder revised the labels. We calculate the inter-coder agreement after this step. We had a final meeting to resolve the disagreement in our labeling results and reached the final label for each method. For each difference in our labels, the two coders and a third author discussed the conflict and reached a consensus.

\subsection{Association rule mining}
\label{subsec:assocaitionRule}
The association rule mining technique has been used in many studies in software engineering research such as cascade library updates, API auto-compilation, and software error prediction.

Association rules are expressed in the form of X $\Rightarrow$ Y. X is called the antecedent and Y is called the consequence (if X then Y). We use this technique to uncover how DL libraries are associated with each other. Every mined rule has a \texttt{support} which is the frequency of the occurrence of an item set X in the transaction database. Also, we define the confidence (conditional probability) of a mined rule as the frequency of X and Y co-occur divided by the frequency of X. In our work, we use the \textbf{\texttt{Appriori}} algorithm  as our association rule technique. This algorithm is used for mining frequent item sets and devising association rules from a transnational database. The \textbf{\texttt{Appriori}} algorithm starts (1) by computing the \texttt{support} of an item set of in the transnational database, then (2) prune the item sets with a support less than a given threshold and finally (3) generate association rules from the above frequent item sets. 

\noindent\begin{minipage}{.5\linewidth}
\begin{equation}
    Support(X) = \frac{frequency(X) }{\textnormal{Total number of transaction}}
\end{equation}
\end{minipage}%
\begin{minipage}{.5\linewidth}
\begin{equation}
    Confidence(X \Rightarrow Y) = \frac{Support(X\cup Y)}{Support(X)}
\end{equation}
\end{minipage}
\section{Results}
\label{sec:results}
In this section, we discuss the results of the research questions (RQs). We employ qualitative and quantitative analysis to answer each question.

\subsection{\RQOne{}}
\label{sec:rq1}

\begin{figure}[!t]
\centering
\captionsetup{justification=centering}
    \centering
\includegraphics[width=1\textwidth]{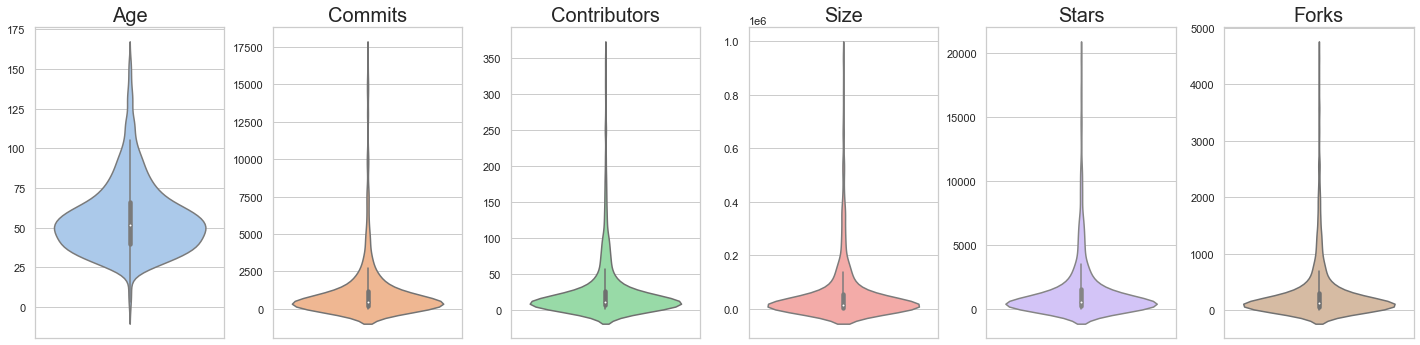}
    \caption{Summary of the projects that uses Deep learning libraries characteristics. We show their total Age in months (i.e., difference between the creation date and the latest update date), number of commits, number of contributors, number of forks, project’s size in term of lines of code, and the number of stars.}
    \label{fig:overview}
\label{fig:repo_desc}
\vspace{-3mm}
\end{figure}
\subsubsection{\textbf{Motivation}} 
Although Deep learning continues to fascinate us with endless possibilities it is still challenging to choose the right deep learning libraries. The purpose of RQ1 is to highlight the most used Deep learning libraries, understand the evolution of the deep learning libraries usage over time and give elaborated information about the projects using these libraries. This insight can help machine learning practitioners to select the appropriate
Deep learning library for their work. Moreover, the information
from the projects using the libraries can help DL libraries maintainers 
improve their tools in future releases.

\subsubsection{\textbf{Approach}} 
To identify the most used deep learning libraries in our corpus $P$. For each library $l \in{L}$ and for version $v \in {V}$ of $l$, we collected the function call in the GitHub projects. Besides, we analyzed the characteristics of the GitHub projects p $\in{P}$. 

\noindent \textbf{Collecting function calls of Deep learning libraries in GitHub projects:} To understand the popularity of the deep learning libraries, for each project $p$, we analyzed the code source files and extracted the calls that invoke the function of each deep learning library(s) $l \in{L}$. We follow the corresponding steps to extract the function calls from the code source files:
\begin{enumerate}
    \item Using the \texttt{nbconvert\footnote{\url{https://nbconvert.readthedocs.io/en/latest/}}} and \texttt{nbformat\footnote{\url{https://nbformat.readthedocs.io/en/latest/}}} python libraries, we convert the notebook files to Python files.
    \item Then, we convert the Python2 source code files to python3.
    \item For each library  $l \in{L}$, we extract the import statement and function call using in each source code files using the abstract syntax tree (AST)\footnote{\url{https://docs.python.org/3/library/ast.html}} python library.
    \item we also extract the functions: \begin{enumerate}
        \item Directly called in the source code file.
        \item called in the right side of an assignment.
        \item called in the condition clause (if else), exception block (try except), called inside a loop (While, for), return statement, assert, data structure (list, tuple) and unary operation statements.
        \item Child functions of a class in which the parent function calls a deep learning library.
    \end{enumerate}
\end{enumerate}

To identify the popularity of the Deep learning libraries, we first for each project $p$ $\in{P}$ we counted the median of total number of calls across all it is version $V$. Then, for library l $\in{L}$ we counted the total number of function calls among all projects. Finally, we sorted the Deep learning based on the total number function calls. 

\noindent \textbf{Analyzing the characteristics of GitHub projects:} To understand the characteristics of the GitHub projects that use Deep learning library(s). For each project p $\in{P}$, we analyzed the information of the projects like the number of contributors, number of forks, size, age, star, and number of commits.

\begin{figure}
\centering
\begin{minipage}{.5\textwidth}
  \centering
  \includegraphics[width=\linewidth]{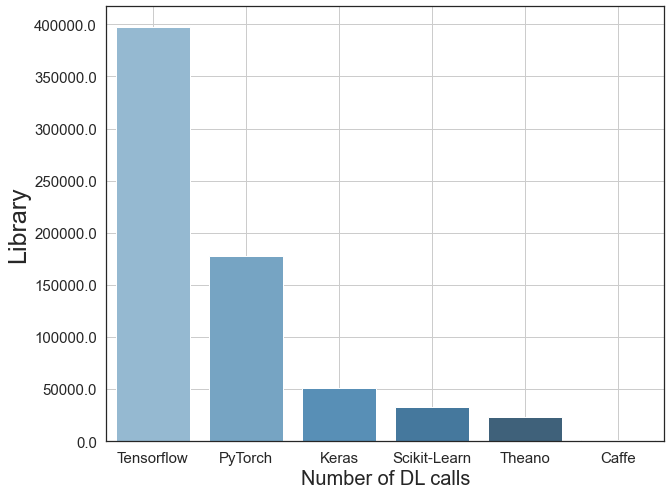}
  \captionof{figure}{Deep libraries function calls}
  \label{fig:DL_popularity}
\end{minipage}%
\begin{minipage}{.5\textwidth}
  \centering
  \includegraphics[width=\linewidth]{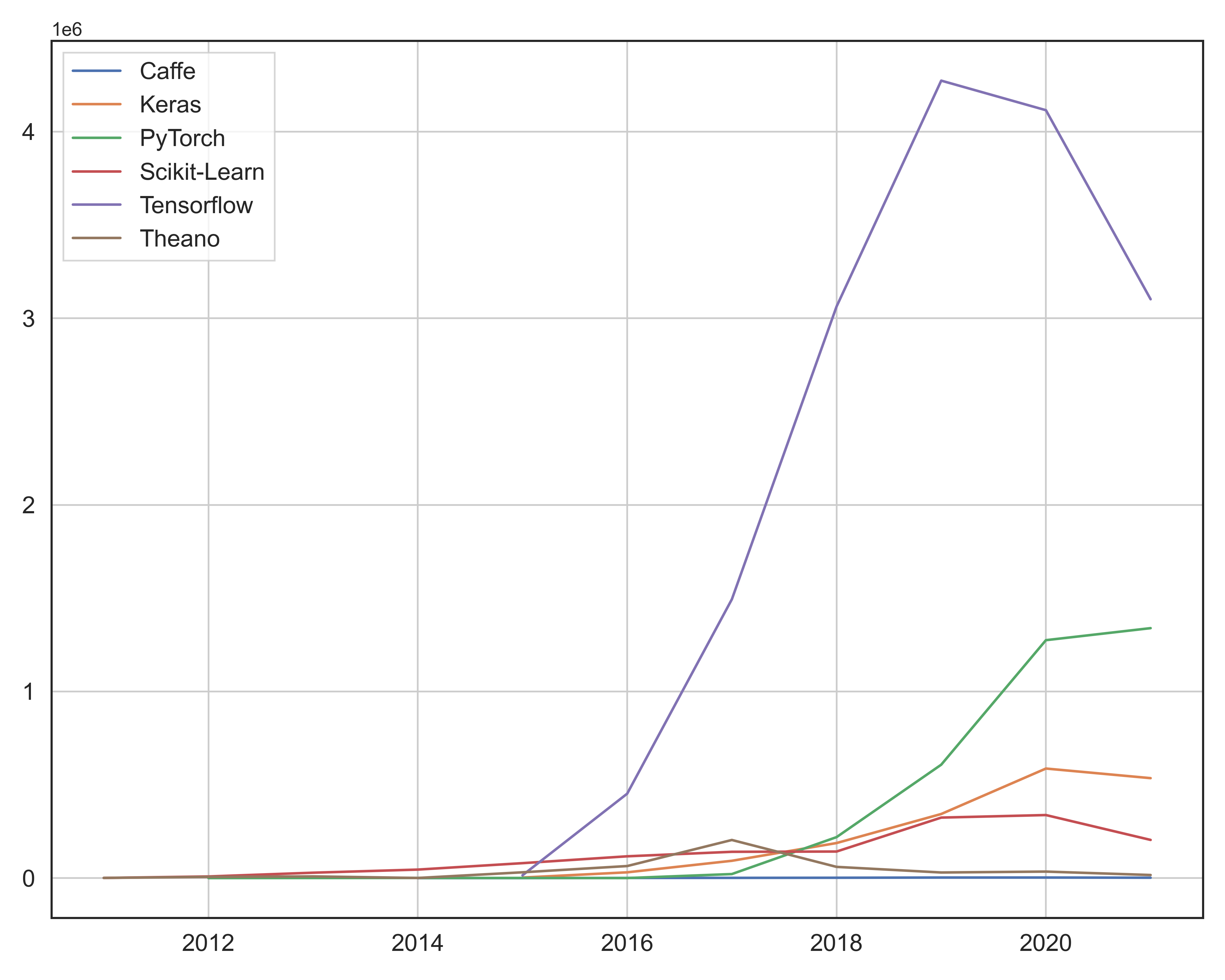}
  \captionof{figure}{Evolution of deep learning libraries popularity overtime}
  \label{fig:DL_overTime}
\end{minipage}
\vspace{-5mm}
\end{figure}

\subsubsection{\textbf{Result}} Fig.~\ref{fig:DL_overTime} shows the total number of function calls per year for each DL library. As seen in Fig.~\ref{fig:DL_overTime}, between 2015 and 2019 \texttt{Tensorflow} show an increasing trend and has the highest number of function calls. However in 2019, the number of function calls is decreasing while the number of function calls of \texttt{PyTorch} is increasing. This observation is correlated with the new release of \texttt{Tensorflow 2.0} in September 2019 which indicate that practitioners find it hard to adapt to \texttt{Tensorflow 2.0}. Also the increase in \texttt{PyTorch} after 2019 highlight that practitioners are moving away from \texttt{Tensorflow} to \texttt{PyTorch}. This calls for \texttt{Tensorflow} library maintainers to review the new releases. Fig.~\ref{fig:DL_popularity} illustrate the total number of deep learning function calls among the projects. As seen in the Fig.~\ref{fig:DL_popularity}, the top 3 deep learning libraries are \texttt{Tensorflow}, \texttt{PyTorch} and \texttt{Keras}. 87\% of the deep learning calls are assigned to these top 3 deep learning libraries. \texttt{Tensorflow} is by far the most used deep learning libraries almost double of the number of the function calls compared to \texttt{PyTorch} which is the second most used DL library in our corpus. 
In Fig.~\ref{fig:repo_desc}, we present the profile of the project that uses deep learning libraries. For presentation purposes, we used the IQR (Interquartile range) to identify the outlier within the last 5\% of data for each measurement. In Fig.~\ref{fig:repo_desc}, we show the 95\% of the data. From Fig.~\ref{fig:repo_desc}, we observe that 75\% of the projects have less than 75 moth old (Q3 is 67 months), with 1378 commits, 29 contributor, 68,027 size (LOC), 350 forks and 1,637 stars. More over, we see that project that uses deep learning libraries have also mature projects with more than 144 months old (12 years), sought after with number of stars up to 6,664, with many contributors as much as 337 and large size of 751,680 LOC (i.e. pymc \footnote{\url{https://github.com/pymc-devs/pymc}}). 

In GitHub projects, forks usually indicate a contribution to the project either by fixing bugs, adding new features or for optimizations~\cite{10.1007/s10664-016-9436-6}. DL project shows a high number of forks which request from library maintainers to invest resources for mitigating the challenges.

\begin{mybox}{RQ1: \RQOne }
    In particular:
    \begin{itemize}[itemsep = 3pt, label=\textbullet, wide = 0pt]
      \item Between 2015 and 2019, DL library usage has grown from 2\% to 80\%.
      \item \texttt{Tensorflow} is the most popular deep learning library, but \texttt{PyTorch} starts to gain popularity since 2019 while \texttt{Tensorflow} popularity is decreasing. This indicate a difficulty for practitioner to adapt to \texttt{Tensorflow 2.0} and start to migrate to \texttt{PyTorch}.
      \item Most of the project that uses DL libraries are popular according to their number of stars but show a high number of forks which suggest that library maintainers need to put more resources to mitigate the challenges that practitioners are facing.
          \end{itemize}
\end{mybox}

\subsection{\RQTwo{}}
\label{sec:rq2}
\subsubsection{\textbf{Motivation}} With too many deep learning libraries sharing the same purpose, developers will feel submerged and confused about choosing the most suited library for their requirements. Also, they need to invest time and effort to become experts in the DL library. Moreover, pioneer companies such as Intel, NVIDIA, Apple, Google, and Facebook are massively investing in hardware to accelerate and optimize the ML workloads in real-time, but without having a view of the combination of the DL libraries used in practice they are left in the dark. In RQ2, we discuss and answer the question What is the degree of dependency between DL libraries? How many DL libraries do developers use in software projects? How are the DL libraries distributed across the ML development workflow? Answering these questions will motivate researchers to understand the challenges of using many DL libraries, and help hardware and library developers build an instructive decision.

\subsubsection{\textbf{Approach}} To identify the most used deep learning  combination in our corpus $P$. For each version $v_{i}$ $\in{V}$ of a project $p$, we analyzed the function call and identify the dependency model in $p$ as defined in section~\ref{subsubsec:DModel}. 

\noindent \textbf{Identifying the dependency model:} As described in section \ref{subsubsec:DLapplication}, in order to identify the DL libraries used in the projects. First, we perform a static code analysis then we look in-depth into the different DL library combinations by inspecting the different imports of DL libraries across all the project releases. In fact if a project p $\in{P}$ in a version $v_{i}$ $\in{V}$ import library $l_{1}$ and $l_{2}$ $\in{L}$ then the used combination which describes the dependency model in the project is a tuple (p,$v_{i}$,$l_{1}$,$l_{2}$). Finding such combinations can help hardware developers and DL library providers improve their products or integrate new features that pushed the users to use a combination of libraries.

\noindent \textbf{Identifying the degree of dependency between libraries}: As described in the section \ref{subsubsec:projectcategories}, to further discover the dependencies between the DL libraries, we have assigned the set of project $P$ into four categories. Tensorflow-based projects, PyTorch-based projects, Keras-based projects, and Scikit-learn-based projects. We analyze the degree of dependency over two dimensions:
\begin{itemize}
    \item \textbf{File degree of dependency}: For each project, we define the file degree of dependency between at least two libraries as the number median number of files with calls of library $l_m$ $\in{L}$ divided by the median number of files of $p$ $\in{P}$ across all his versions $V$. 
    \item \textbf{Function degree of dependency}: For each project, we define the function degree of dependency between at least two libraries as the number median number of functions with calls of library $l_m$ $\in{L}$ divided by the median number of function in p $\in{P}$ across all his versions $V$.
\end{itemize}
    
\subsubsection{\textbf{Result:}} 
\begin{table*}[h]
    \begin{minipage}{.5\linewidth}
      \caption{Library usage in Deep Learning projects}
    \centering
    \scalebox{0.8}{
        \begin{tabular}{@{}ll|ll@{}}
        \toprule
        \multicolumn{2}{c|}{\cellcolor[HTML]{EFEFEF}\begin{tabular}[c]{@{}c@{}}Single DL library projects\\ (823 Projects)\end{tabular}} &
          \multicolumn{2}{c}{\cellcolor[HTML]{EFEFEF}\begin{tabular}[c]{@{}c@{}}Multiple DL library projects\\ (661 Projects)\end{tabular}} \\ \midrule
        Library      & \% Projects & Library        & \% Projects       \\ \midrule
        Scikit-Learn &  \Chart{0.57} & Tensorflow $\in{L}$   & \Chart{0.70}      \\
        Keras        &  \Chart{0.48} & Scikit-Learn $\in{L}$ & \Chart{0.68}      \\
        PyTorch      &  \Chart{0.35} & Keras $\in{L}$        & \Chart{0.55}      \\
        Tensorflow   &  \Chart{0.24} & PyTorch $\in{L}$      & \Chart{0.52}      \\
        Caffe        &  \Chart{0.07} & Caffe $\in{L}$        & \Chart{0.06}      \\
        Theano       &  \Chart{0.02} & Theano $\in{L}$       & \Chart{0.05}      \\ \bottomrule
        \multicolumn{3}{l}{L is the set of DL libraries used by a project.}
        \end{tabular}
    }
    \label{table:LibraryUsage}
    \end{minipage}%
    \begin{minipage}{.5\linewidth}
      \caption{Top 9 used DL Library combinations }
      \vspace{-2mm}
    \centering
    \scalebox{0.8}{
        \begin{tabular}{@{}ll@{}}
            \toprule
            \rowcolor[HTML]{EFEFEF} 
            Library Combination                      & \% Projects \\ \midrule
            PyTorch, Scikit-Learn                    & \Chart{0.18}        \\
            Keras, Tensorflow                        & \Chart{0.14}        \\
            Keras, Scikit-Learn, Tensorflow          & \Chart{0.13}        \\
            Keras, PyTorch, Sickit-Learn, Tensorflow & \Chart{0.10}        \\
            Tensorflow, Scikit-Learn                 & \Chart{0.07}        \\
            Tensorflow, Pytorch, Scikit-Learn        & \Chart{0.06}        \\
            Tensorflow, Pytorch                      & \Chart{0.06}        \\
            Keras, Scickit-Learn                     & \Chart{0.06}        \\ 
            keras,tensorflow,torch                   & \Chart{0.06}        \\\bottomrule
        \end{tabular}
    }
    \label{table:LibraryCombination}
    \end{minipage} 
\end{table*}
\begin{table*}[h]
\caption{Library combination used in Tensorflow, PyTorch, Keras and Scikit-learn based projects. We show top 5 most occurring combinations used in the same function per project. The occurrences represent the number of the project that have at least one function that uses a combination of DL libraries.}
\scalebox{0.50}{
\begin{tabular}{@{}cccc|cccc|cccc|cccc@{}}
\toprule
\multicolumn{4}{c|}{\cellcolor[HTML]{EFEFEF}\begin{tabular}[c]{@{}c@{}}PyTorch \\ (140)\end{tabular}} & \multicolumn{4}{c|}{\cellcolor[HTML]{EFEFEF}\begin{tabular}[c]{@{}c@{}}Tensorflow\\ (185 projects)\end{tabular}} & \multicolumn{4}{c|}{\cellcolor[HTML]{EFEFEF}\begin{tabular}[c]{@{}c@{}}Keras\\ (89 Projects)\end{tabular}} & \multicolumn{4}{c}{\cellcolor[HTML]{EFEFEF}\begin{tabular}[c]{@{}c@{}}Scikit-learn \\ (26 projects)\end{tabular}} \\ \midrule
\multicolumn{1}{l}{Tensorflow} & \multicolumn{1}{l}{Keras} & \multicolumn{1}{l|}{Scikit-learn} & \multicolumn{1}{l|}{Occurence} & \multicolumn{1}{l}{PyTorch} & \multicolumn{1}{l}{Keras} & \multicolumn{1}{l|}{Scikit-learn} & \multicolumn{1}{l|}{Occurence} & \multicolumn{1}{l}{PyTorch} & \multicolumn{1}{l}{Tensorflow} & \multicolumn{1}{l|}{Scikit-learn} & \multicolumn{1}{l|}{Occurence} & PyTorch & Tensorflow & \multicolumn{1}{c|}{Keras} & Occurence \\ \midrule
 &  & \multicolumn{1}{c|}{\checkmark} & 82 &  & \checkmark & \multicolumn{1}{c|}{} & 137 &  & \checkmark & \multicolumn{1}{c|}{} & 71 &  &  & \multicolumn{1}{c|}{\checkmark} & 14 \\
\checkmark &  & \multicolumn{1}{c|}{} & 48 & \checkmark &  & \multicolumn{1}{c|}{} & 52 &  &  & \multicolumn{1}{c|}{\checkmark} & 33 & \checkmark &  & \multicolumn{1}{c|}{} & 8 \\
\checkmark & \checkmark & \multicolumn{1}{c|}{} & 47 & \checkmark & \checkmark & \multicolumn{1}{c|}{} & 50 &  & \checkmark & \multicolumn{1}{c|}{\checkmark} & 10 &  & \checkmark & \multicolumn{1}{c|}{} & 4 \\
\checkmark &  & \multicolumn{1}{c|}{-} & 8 &  &  & \multicolumn{1}{c|}{\checkmark} & 44 & \checkmark & \checkmark & \multicolumn{1}{c|}{} & 7 &  & \checkmark & \multicolumn{1}{c|}{\checkmark} & 3 \\
\checkmark & \checkmark & \multicolumn{1}{c|}{\checkmark} & 5 &  & \checkmark & \multicolumn{1}{c|}{\checkmark} & 15 & \checkmark & \checkmark & \multicolumn{1}{c|}{} & 7 &  &  & \multicolumn{1}{c|}{} &  \\ \bottomrule
\end{tabular}
}
\label{tab:combination_lib}
\end{table*}

\subsubsubsection{\textbf{Popular dependencies of DL libraries}}
\textbf{From table~\ref{table:LibraryUsage}, we highlight that 823 (55.4\%) of the selected project use a single DL library (e.g., Tensorflow or PyTorch), while the rest of our project 661 (45.6\%) use different combinations of DL libraries.} Table~\ref{table:LibraryUsage} illustrate that, while \texttt{Scikit-Learn} is the most independent library with 57\% of 823 projects using a single library, \texttt{Scikit-Learn}, \texttt{Tensorflow}, \texttt{Keras} and \texttt{PyTorch} are the most popular libraries used in combination with another DL library. For instance, this questions the completeness and correctness of some of the DL libraries' functionalities and calls for libraries maintainers to invest further resources in future releases.

\noindent We found 37 different DL library combinations, but for presentation purposes in Table \ref{table:LibraryCombination} we only illustrate the top 9 most occurring ones. 
From table \ref{table:LibraryCombination}, we observe that users opt to use a different combination of DL libraries. In fact, during our study, we have found that developers use from 2 to 4 various libraries simultaneously during the life cycle of a project. 
\textit{PyTorch} and \textit{Scikit-Learn} is the most used combination representing 18\% of the studied projects while \textit{Keras} and \textit{Tensorflow} is the second most popular combination with 14\% of the projects. Also, we observe that developers have used 4 different DL libraries in 10\% of the projects shaped in \textit{Keras}, \textit{PyTorch}, \textit{Scikit-Learn}, and \textit{Tensorflow} combination. Moreover, one of the most intriguing combinations that we observe is between \textit{PyTorch} and \textit{Tensorflow} with 22\% of the studied project showing a trace of using the two DL libraries. 

\subsubsubsection{\textbf{Dependency relation between DL libraries}}
In table~\ref{tab:combination_lib}, we represent the dependency between each library $l$ and the rest of the studied DL libraries. We found that \texttt{PyTorch} projects show high dependency with \texttt{Tensorflow}. In fact, from table~\ref{tab:combination_lib} \texttt{TensorFlow} is used in \texttt{PyTorch} projects in almost all top 5 combinations except with \texttt{Scikit$-$learn}. \texttt{TensorFlow} projects are mostly dependent of \text{Keras} with 137 projects of 185 uses the combination Tensorflow $\Leftrightarrow$ Keras in the same function. In \texttt{Tensorflow} projects, the combination Tensorflow $\Leftrightarrow$ PyTorch is the  second most frequent with 102 projects using this combination in the same function while the combination Tensorflow $\Leftrightarrow$ Scikit$-$learn is the least used one which is present in 60 projects. Keras projects are highly dependent on Tensorflow. As a matter of a fact, Keras and Tensorflow are present in all different combinations except within the combination Keras $\Leftrightarrow$ Scikit$-$learn. Besides, \texttt{Keras} shows a low dependency with \texttt{PyTorch} where the combination Keras $\Leftrightarrow$ PyTorch is used in the same function only in 7 projects. Finally, \texttt{Scikit$-$learn} projects are the least dependent on other DL libraries with only 26 projects using multiple DL libraries in the same function. Scikit$-$learn $\Leftrightarrow$ Keras appears the most frequently with only 14 projects using this combination. The combination Tensorflow $\Leftrightarrow$ PyTorch is omnipresent in both \texttt{TensorFlow} and \texttt{PyTorch} projects. This observation suggests a limitation of the two libraries and shows that the two libraries are complementary for practitioners. This observation first suggests that hardware developers should not focus on optimizing their hardware for one specific DL library~\cite{intel}~\cite{IBM2}~\cite{Apple}, and second library maintainers can use this observation to improve their future versions and realize more complete libraries.

\begin{figure}[!t]
\centering
\captionsetup{justification=centering}
    \centering
\includegraphics[width=0.9\textwidth]{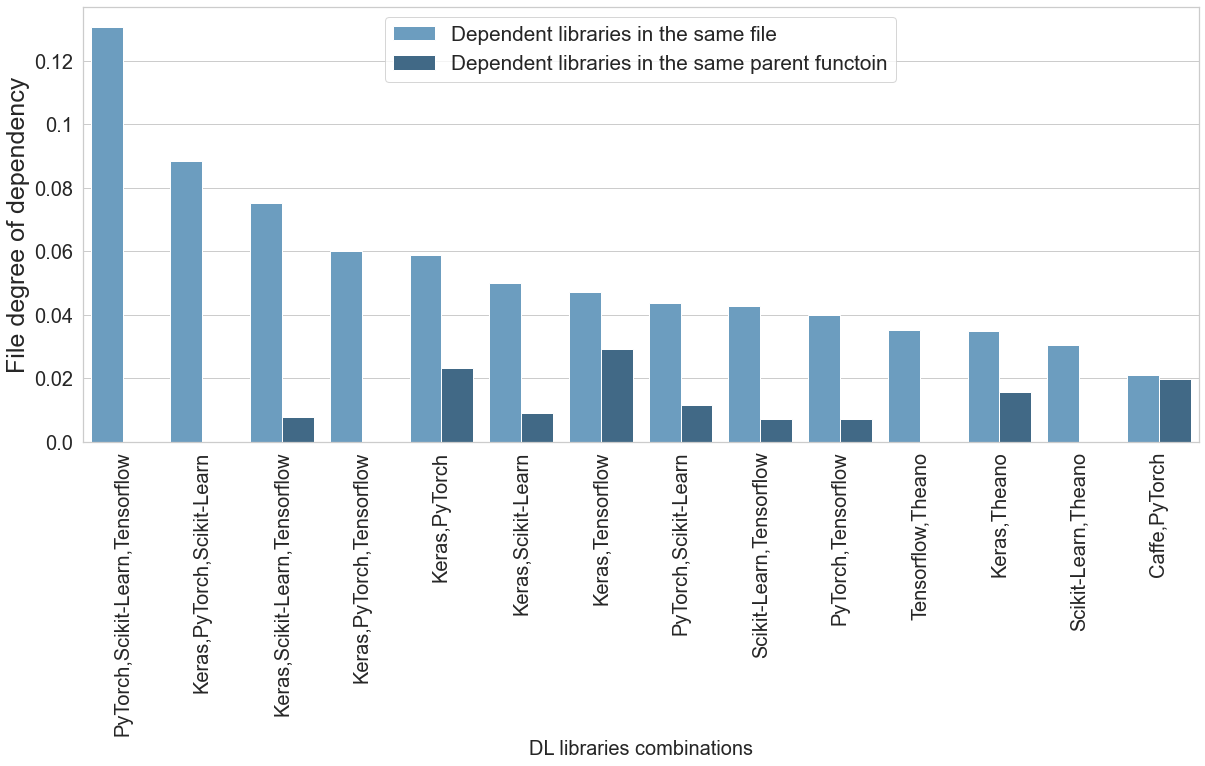}
    \caption{Distribution of the degree of the dependencies in the project's latest version}
    \label{fig:overview}
\label{figure:DistributionDegreeDep}
\vspace{-5mm}
\end{figure}

\subsubsubsection{\textbf{Degree of dependencies between DL libraries}}
Fig.~\ref{figure:DistributionDegreeDep} illustrates the distribution of the file degree of the dependencies across the projects. From the list of project $P$, we detect 705 projects using multiple DL libraries in the same file in which we select 338 projects that have at least 20 median numbers of files across the versions $V$. \textbf{The four most combination with the highest file degree dependency, use three different deep learning libraries.} From Fig.~\ref{figure:DistributionDegreeDep}, \texttt{Scikit$-$Learn} is present in the first three combination. Developers tend to use \texttt{Scikit$-$Learn} for data preparation and processing while using \texttt{PyTorch}, \texttt{Keras} and \texttt{Tensorflow} for \texttt{training} and \texttt{testing} the model. The libraries \texttt{PyTorch} and\texttt{Tensorflow} show the highest file degree dependency with 12\%. Following \texttt{PyTorch}, \texttt{Keras} are the second most dependent libraries with 8\% file degree of dependency. Finally, The libraries \texttt{Keras} and \texttt{Tensorflow} have the third highest degree of dependencies with at least 7\% number of files in the studied project using this combination as shown in the Fig. \ref{figure:DistributionDegreeDep}. We explain this by the fact that \texttt{Keras} is a high-level API of \texttt{TensorFlow 2.0} and acts as an interface for \texttt{Tensorflow}.

Fig.~\ref{figure:DistributionDegreeDep} illustrates the distribution of the function degree of the dependencies across the projects. From the list of project P, we detect 449 projects using multiple DL libraries in the same function in which we select 229 projects that have at least 70 median total number of functions across the versions $V$. 

\noindent \textbf{The libraries \texttt{Keras} and \texttt{Tensorflow} have the highest function degree of dependencies with almost 3\% number of functions in the studied project using this combination as shown in the Fig.~\ref{figure:DistributionDegreeDep}}. We explain this by the fact that \texttt{Keras} is a high-level API of \texttt{TensorFlow 2.0} and acts as an interface for \texttt{Tensorflow}. Moreover, we observe that \texttt{Keras} and \texttt{PyTorch} combination have the second highest file degree of dependency with more than 2\%.   

From Fig.~\ref{figure:DistributionDegreeDep} we conclude that \texttt{keras and Pytorch} show a high level of dependency in both  the same function and the same file, similarly to \texttt{Keras} and \texttt{Tensorflow} libraries.

\begin{mybox}{}

    In particular:
    \begin{itemize}[itemsep = 3pt, label=\textbullet, wide = 0pt]

      \item Practitioners tend to use a combination of 2 and 3 libraries during the life cycle of the project.
          \item Practitioners tend to use the combination of \texttt{PyTorch} and \texttt{Scikit-learn} and, \texttt{Tensorflow} and \texttt{Keras} the most.
        \item \texttt{Sickit-learn} is the most dependent library to the rest of the DL libraries.
      \item \texttt{PyTorch} show a high dependency with \texttt{Tensorflow}, while \texttt{Tensorflow} shows a high dependency with \texttt{Keras} and \texttt{PyTorch}.
      \item \texttt{Keras and Pytorch} have a high dependency degree in the file level and function level with respective value of 13\% and 2.5\%.
    \end{itemize}
\end{mybox}

\subsection{\RQThree{}}
\label{sec:rq3}
\subsubsection{\textbf{Motivation}} Although all deep learning libraries provide similar functionalities to practitioners, some of them are better suited for a certain stage of a machine learning software development pipeline. DL practitioners may use DL libraries for different reasons across the machine learning pipeline. We look in-depth into the DL library(s) across the different stages of the machine learning pipeline and we analyze the dependency of the DL libraries across the different stages. RQ2 is going to help DL library(s) provider(s) understand the usage of their respective tool and give them the hand to improve future releases in a more effective way. 

Understanding the DL library dependency across the ML pipelines can give the providers insight into their libraries' limitations, and on the other hand, it can help practitioners better exploit the DL libraries in different ML stages.

\subsubsection{\textbf{Approach}} To understand the dependency between DL libraries across the ML development pipeline, we label the extracted function calls in Section~\ref{subsec:codeanalysis}, uncover the dependency between the libraries, and apply the association rule algorithm discussed in Section~\ref{subsec:assocaitionRule} to disclose any patterns in the usage of the DL libraries by practitioners. We focused on the function calls that are commonly used. To identify the function calls, we compute the normalized Shannon entropy value and keep a function that has a value above 30\%. In the following, first, we explain the process of selecting function calls based on the normalized Shannon entropy. The second is the process of manual labeling.

\noindent \textbf{Selecting function calls based on Shannon entropy:} To compute the distribution of the function calls across the projects, we used the normalized Shannon entropy~\cite{10.1007/s11219-017-9361-y}~\cite{10.1145/584091.584093}.

\noindent\begin{minipage}{.3\linewidth}
\begin{equation}
\label{eqn:Shanon}
    \mathit{H_{n}}\left ( \mathit{f} \right )= -\sum_{i=1}^{i=n} p_{i}*log_{n}\left( p_{i}\right )
\end{equation}
\end{minipage}%
\begin{minipage}{.7\linewidth}
\begin{equation}
    p_{i}= \frac{\text{Total number of occurrence of a function call in a project $i$}}{\text{Total number of occurrence of function $f$ in all the projects}}
\end{equation}
\end{minipage}

Let $f$ a function calls $\in$ to the set of function calls $F$, $H$ the equation~\ref{eqn:Shanon} of normalized Shannon entropy, $n$ the number of unique project, $i$ the number of the the unique project and $p_{i}$ is the probability to use the function $f$ in a project $i$ where $p_{i} > 0$, and $\sum_{i=1}^{i=n} p_{i} = 1$. If $\mathit{H_{n}}\left ( \mathit{f} \right ) = 1$, $f$ is maximal which indicate each project $p_{i} \in{P}$ have the same probability using the function $f$. Hence the highest Shannon entropy value indicates that a function $f$ is more likely used by a practitioner compared to function with low Shannon entropy value.  

The function $f$ that has a $\mathit{H_{n}\left ( \mathit{f} \right )>= 30\%}$ were choosing for manual labeling and association rule mining. In the rest of the paper, we refer to filtered data as $D_{s}$. 

\noindent \textbf{Manual labeling:} In this section, we are looking to understand the different DL libraries combination across the ML pipeline. Accordingly for each library $l$ $\in{L}$ we sample a representative sample with a 99\% confidence level and a margin error of 5\% from the list of functions of library $l$ in $D_{s}$. Two authors assigned two labels for each DL function call: (1) Data, model, framework, or utility related; and (2) ML development workflow stage. For the labeling process, we used the approach described in the manual analyses section~\ref{subsec:manualanalysis}. 

\noindent \textbf{Pattern recognition:} In this section, we are looking to extract hidden patterns used by practitioners in DL projects. Using the association rules algorithms explained in section~\ref{subsec:assocaitionRule}, we extract the used DL libraries functions rules by practitioners in the different machine learning pipelines. Accordingly, for each set of function calls used in the same machine learning pipeline, we apply the association rules algorithms, evaluate the confidence and support of the rules and report the most relevant ones. To understand the reasons behind using multiple deep learning libraries, and to uncover hidden patterns in the dependency of DL libraries, we report the most prevalent combination between \texttt{Keras}$\Leftrightarrow$\texttt{TensorFlow}, \texttt{PyTorch}$\Leftrightarrow$\texttt{TensorFlow}, \texttt{Keras}$\Leftrightarrow$\texttt{TensorFlow}, \texttt{Scikit$-$learn}$\Leftrightarrow$\texttt{TensorFlow} and \texttt{Scikit$-$learn}$\Leftrightarrow$\texttt{PyTorch}.  
\begin{table*}[h]
\caption{Summary of the labeled calls from data set $D_{s}$ for each DL library }
\scalebox{0.45}{
    \begin{tabular}{lccccccccccccc}
    \cline{2-14}
     & \multicolumn{3}{c|}{\textbf{Date  processing related}} & \multicolumn{3}{c|}{\textbf{Model related}} & \multicolumn{6}{c|}{\textbf{Framework related}} & \multicolumn{1}{l}{\multirow{2}{*}{\textbf{Utility related}}} \\ \cline{2-13}
     & \multicolumn{1}{c|}{\textbf{\begin{tabular}[c]{@{}c@{}}Data \\ collection\end{tabular}}} & \multicolumn{1}{c|}{\textbf{\begin{tabular}[c]{@{}c@{}}Data\\ Cleaning\end{tabular}}} & \multicolumn{1}{c|}{\textbf{\begin{tabular}[c]{@{}c@{}}Feature \\ engeneering\end{tabular}}} & \multicolumn{1}{c|}{\textbf{\begin{tabular}[c]{@{}c@{}}Model\\ training\end{tabular}}} & \multicolumn{1}{c|}{\textbf{\begin{tabular}[c]{@{}c@{}}Model \\ evaluation\end{tabular}}} & \multicolumn{1}{c|}{\textbf{\begin{tabular}[c]{@{}c@{}}Distributed \\ computation\end{tabular}}} & \multicolumn{1}{c|}{\textbf{\begin{tabular}[c]{@{}c@{}}Mathematical \\ calculation\end{tabular}}} & \multicolumn{1}{c|}{\textbf{\begin{tabular}[c]{@{}c@{}}Tensor and array \\ manupilation\end{tabular}}} & \multicolumn{1}{c|}{\textbf{configuration}} & \multicolumn{1}{c|}{\textbf{Testing}} & \multicolumn{1}{c|}{\textbf{Logging}} & \multicolumn{1}{c|}{\textbf{Debugging}} & \multicolumn{1}{l}{} \\ \hline
    \multicolumn{1}{l|}{\textbf{TensorFlow}} & 8 & 30 & 51 & 175 & 49 & 15 & 91 & 57 & 22 & 12 & 11 & 1 & 72 \\
    \multicolumn{1}{l|}{\textbf{PyTorch}} & 9 & 24 & 71 & 231 & 11 & 14 & 64 & 46 & 0 & 0 & 10 & 0 & 48 \\
    \multicolumn{1}{l|}{\textbf{Keras}} & 2 & 22 & 43 & 239 & 19 & 29 & 38 & 27 & 0 & 0 & 0 & 0 & 23 \\
    \multicolumn{1}{l|}{\textbf{Sickit-learn}} & 9 & 41 & 119 & 153 & 112 & 2 & 14 & 9 & 3 & 20 & 0 & 0 & 37 \\
    \multicolumn{1}{l|}{\textbf{Caffe}} & 0 & 10 & 19 & 87 & 5 & 17 & 5 & 3 & 0 & 2 & 1 & 0 & 38 \\ \hline
    \end{tabular}
}
\vspace{-3mm}
\end{table*}

\subsubsubsection{\textbf{Distribution of DL libraries dependency across ML development workflow}}
\label{subsubsec:RQ3}
\textbf{From table~\ref{tab:dataProcessing}, we highlight first a frequent dependency between \texttt{Keras}$\Leftrightarrow$\texttt{TensorFlow} and \texttt{Scikit$-$learn} $\Leftrightarrow$ \texttt{PyTorch} in the \texttt{data collection} stage, second a rare dependency between \texttt{Keras} $\Leftrightarrow$ \texttt{TensorFlow} in the data cleaning stage and finally a recurrent dependency between \texttt{PyTorch} $\Leftrightarrow$ \texttt{TensorFlow} and \texttt{Scikit$-$learn} $\Leftrightarrow$ \texttt{PyTorch} in the Feature engineering stage.} Table~\ref{tab:dataProcessing} illustrates the five most frequent dependencies between the deep learning libraries across the 3 stages, \texttt{data collection}, \texttt{data cleaning}, and \texttt{feature engineering}. In fact, we observe that the combination \texttt{Keras} $\Leftrightarrow$ \texttt{TensorFlow} is the most frequent in the three stages with 7\% in the data collection stage, 8\% in the feature engineering stage and a rare presence of 0.5\% in the data cleaning stage. Also, the dependency between \texttt{Scikit-learn} $\Leftrightarrow$ \texttt{TensorFlow} is present in the data collection and Feature engineering stage with respective values of 7\% and 10\% of the project. Besides, in the data-processing category, we detect a recurrent combination between \texttt{PyTorch} $\Leftrightarrow$ \texttt{TensorFlow} among the three different stages. We show 12\% of the projects in the feature engineering stage, 6\% of the projects in the data collection stage, and a low frequency in the data cleaning stage with only 0.1\%. Finally, from Table~\ref{tab:dataProcessing}, we see that 3\% of the project in the data collection stage and 4\% of the project in the feature engineering stage show a dependency between three DL libraries \texttt{PyTorch} $\Leftrightarrow$ \texttt{TensorFlow} $\Leftrightarrow$ \texttt{Keras}. This can be explained by the fact that some projects are in the process of migrating partially the DL libraries from one to another.

\begin{table*}[h]
    \begin{minipage}{.4\linewidth}
      \caption{Library dependency in Data processing category}
    \centering
    \scalebox{0.5}{
        \begin{tabular}{@{}ccccc|cccc|ccccc@{}}
            \hline
            \rowcolor[HTML]{EFEFEF} 
            \multicolumn{5}{c|}{\cellcolor[HTML]{EFEFEF}\textbf{Data collection}} & \multicolumn{4}{c|}{\cellcolor[HTML]{EFEFEF}\textbf{Data Cleaning}} & \multicolumn{5}{c}{\cellcolor[HTML]{EFEFEF}\textbf{Feature  engineering}} \\ \hline
            \rowcolor[HTML]{EFEFEF} 
            \multicolumn{1}{l}{\cellcolor[HTML]{EFEFEF}\rotatebox[origin=c]{90}{\textbf{Keras}}} & \multicolumn{1}{l}{\cellcolor[HTML]{EFEFEF}\rotatebox[origin=c]{90}{\textbf{Scikit$-$learn}}} & \multicolumn{1}{l}{\cellcolor[HTML]{EFEFEF}\rotatebox[origin=c]{90}{\textbf{TensorFlow}}} & \multicolumn{1}{l|}{\cellcolor[HTML]{EFEFEF}\rotatebox[origin=c]{90}{\textbf{PyTorch}}} & \rotatebox[origin=c]{90}{\textbf{\% projects}} & \multicolumn{1}{l}{\cellcolor[HTML]{EFEFEF}\rotatebox[origin=c]{90}{\textbf{Keras}}} & \multicolumn{1}{l}{\cellcolor[HTML]{EFEFEF}\rotatebox[origin=c]{90}{\textbf{PyTorch}}} & \multicolumn{1}{l|}{\cellcolor[HTML]{EFEFEF}\rotatebox[origin=c]{90}{\textbf{TensorFlow}}} & \rotatebox[origin=c]{90}{\textbf{\% projects}} & \multicolumn{1}{l}{\cellcolor[HTML]{EFEFEF}\rotatebox[origin=c]{90}{\textbf{Keras}}} & \multicolumn{1}{l}{\cellcolor[HTML]{EFEFEF}\rotatebox[origin=c]{90}{\textbf{Scikit$-$learn}}} & \multicolumn{1}{l}{\cellcolor[HTML]{EFEFEF}\rotatebox[origin=c]{90}{\textbf{TensorFlow}}} & \multicolumn{1}{l|}{\cellcolor[HTML]{EFEFEF}\rotatebox[origin=c]{90}{\textbf{PyTorch}}} & \rotatebox[origin=c]{90}{\textbf{\% projects}} \\ \midrule

            \checkmark &  & \checkmark & \multicolumn{1}{c|}{} & 7\% & \checkmark &  & \multicolumn{1}{c|}{\checkmark} & 0,5\% &  &  & \checkmark & \multicolumn{1}{c|}{\checkmark} & 12\% \\ \midrule
             & \checkmark &  & \multicolumn{1}{c|}{\checkmark} & 7\% &  & \checkmark & \multicolumn{1}{c|}{\checkmark} & 0,1\% &  & \checkmark &  & \multicolumn{1}{c|}{\checkmark} & 10\% \\ \midrule
             &  & \checkmark & \multicolumn{1}{c|}{\checkmark} & 6\% &  &  & \multicolumn{1}{c|}{} &  & \checkmark &  & \checkmark & \multicolumn{1}{c|}{} & 8\% \\ \midrule
             & \checkmark & \checkmark & \multicolumn{1}{c|}{} & 3\% &  &  & \multicolumn{1}{c|}{} &  & \checkmark &  & \checkmark & \multicolumn{1}{c|}{\checkmark} & 4\% \\ \midrule
            \checkmark &  & \checkmark & \multicolumn{1}{c|}{\checkmark} & 2\% &  &  & \multicolumn{1}{c|}{} &  &  & \checkmark & \checkmark & \multicolumn{1}{c|}{} & 3\% \\ \midrule
        \end{tabular}
    }
    \label{tab:dataProcessing}
    \end{minipage}%
    \begin{minipage}{.6\linewidth}
      \caption{Library dependency in Model related category}
      \vspace{+4mm}
    \centering
    \scalebox{0.5}{
    \begin{tabular}{ccccc|ccccc|cccccc|cccccc}
        \hline
        \rowcolor[HTML]{EFEFEF} 
        \multicolumn{5}{c|}{\cellcolor[HTML]{EFEFEF}\textbf{Model evaluation}} & \multicolumn{5}{c|}{\cellcolor[HTML]{EFEFEF}\textbf{Model training}} & \multicolumn{6}{c|}{\cellcolor[HTML]{EFEFEF}\textbf{Model deployment}} & \multicolumn{6}{c}{\cellcolor[HTML]{EFEFEF}\textbf{Distributed computation}} \\ \hline
        \rowcolor[HTML]{EFEFEF} 
        \multicolumn{1}{l}{\cellcolor[HTML]{EFEFEF}\rotatebox[origin=c]{90}{\textbf{Keras}}} & \multicolumn{1}{l}{\cellcolor[HTML]{EFEFEF}\rotatebox[origin=c]{90}{\textbf{Scikit$-$learn}}} & \multicolumn{1}{l}{\cellcolor[HTML]{EFEFEF}\rotatebox[origin=c]{90}{\textbf{TensorFlow}}} & \multicolumn{1}{c|}{\cellcolor[HTML]{EFEFEF}\rotatebox[origin=c]{90}{\textbf{PyTorch}}} & \rotatebox[origin=c]{90}{\textbf{\% projects}} & 
        \multicolumn{1}{l}{\cellcolor[HTML]{EFEFEF}\rotatebox[origin=c]{90}{\textbf{Keras}}} & \multicolumn{1}{l}{\cellcolor[HTML]{EFEFEF}\rotatebox[origin=c]{90}{\textbf{PyTorch}}} & \multicolumn{1}{l}{\cellcolor[HTML]{EFEFEF}\rotatebox[origin=c]{90}{\textbf{Scikit$-$learn}}} & \multicolumn{1}{c|}{\cellcolor[HTML]{EFEFEF}\rotatebox[origin=c]{90}{\textbf{TensorFlow}}} & \rotatebox[origin=c]{90}{\textbf{\% projects}} & 
        \multicolumn{1}{l}{\cellcolor[HTML]{EFEFEF}\rotatebox[origin=c]{90}{\textbf{Keras}}} & \multicolumn{1}{l}{\cellcolor[HTML]{EFEFEF}\rotatebox[origin=c]{90}{\textbf{Scikit$-$learn}}} & \multicolumn{1}{l}{\cellcolor[HTML]{EFEFEF}\rotatebox[origin=c]{90}{\textbf{TensorFlow}}} & \multicolumn{1}{l}{\cellcolor[HTML]{EFEFEF}\rotatebox[origin=c]{90}{\textbf{Caffe}}} & \multicolumn{1}{c|}{\cellcolor[HTML]{EFEFEF}\rotatebox[origin=c]{90}{\textbf{PyTorch}}} & \rotatebox[origin=c]{90}{\textbf{\% projects}} & 
        \multicolumn{1}{l}{\cellcolor[HTML]{EFEFEF}\rotatebox[origin=c]{90}{\textbf{Keras}}} & \multicolumn{1}{l}{\cellcolor[HTML]{EFEFEF}\rotatebox[origin=c]{90}{\textbf{Scikit$-$learn}}} & \multicolumn{1}{l}{\cellcolor[HTML]{EFEFEF}\rotatebox[origin=c]{90}{\textbf{TensorFlow}}} & \multicolumn{1}{l}{\cellcolor[HTML]{EFEFEF}\rotatebox[origin=c]{90}{\textbf{Caffe}}} & \multicolumn{1}{c|}{\cellcolor[HTML]{EFEFEF}\rotatebox[origin=c]{90}{\textbf{PyTorch}}} & \rotatebox[origin=c]{90}{\textbf{\% projects}} \\ \midrule
        
        \checkmark &  & \checkmark & \multicolumn{1}{c|}{} & 6\% & \checkmark &  &  & \multicolumn{1}{c|}{\checkmark} & 33\% &  &  & \checkmark &  & \multicolumn{1}{c|}{\checkmark} & 11\% &  &  & \checkmark &  & \multicolumn{1}{c|}{\checkmark} & 4\% \\ \midrule
         & \checkmark & \checkmark & \multicolumn{1}{c|}{} & 5\% & \checkmark & \checkmark &  & \multicolumn{1}{c|}{\checkmark} & 12\% & \checkmark &  & \checkmark &  & \multicolumn{1}{c|}{} & 6\% & \checkmark &  & \checkmark &  & \multicolumn{1}{c|}{} & 3\% \\ \midrule
         & \checkmark &  & \multicolumn{1}{c|}{\checkmark} & 4\% &  & \checkmark &  & \multicolumn{1}{c|}{\checkmark} & 12\% & \checkmark &  & \checkmark &  & \multicolumn{1}{c|}{\checkmark} & 4\% & \checkmark &  &  &  & \multicolumn{1}{c|}{\checkmark} & 0.6\% \\ \midrule
        \checkmark & \checkmark &  & \multicolumn{1}{c|}{} & 1\% &  & \checkmark & \checkmark & \multicolumn{1}{c|}{} & 9\% & \checkmark &  &  &  & \multicolumn{1}{c|}{\checkmark} & 4\% &  &  & \checkmark & \checkmark & \multicolumn{1}{c|}{\checkmark} & 0.3\% \\ \midrule
        \checkmark & \checkmark &  & \multicolumn{1}{c|}{\checkmark} & 0.01\% &  &  & \checkmark & \multicolumn{1}{c|}{\checkmark} & 8\% &  &  &  & \checkmark & \multicolumn{1}{c|}{\checkmark} & 2\% &  &  & \checkmark & \checkmark & \multicolumn{1}{c|}{} & 0.3\% \\ \midrule
    \end{tabular}
    }
    \label{tab:modelRelated}
    \end{minipage} 
\end{table*}

\begin{table*}[]
\caption{Library dependency in Framework related category and utility related category}
    \centering
    \scalebox{0.75}{
        \begin{tabular}{cccc|cccc|cccc|cccc|ccccc|cccc}
        \hline
        \rowcolor[HTML]{EFEFEF} 
        \multicolumn{4}{c|}{\cellcolor[HTML]{EFEFEF}\textbf{\begin{tabular}[c]{@{}c@{}}Mathematical   \\ calculation\end{tabular}}} & \multicolumn{4}{c|}{\cellcolor[HTML]{EFEFEF}\textbf{\begin{tabular}[c]{@{}c@{}}Data type and  \\ structure\end{tabular}}} & \multicolumn{4}{c|}{\cellcolor[HTML]{EFEFEF}\textbf{\begin{tabular}[c]{@{}c@{}}Tensor and array \\ manupilation\end{tabular}}} & \multicolumn{4}{c|}{\cellcolor[HTML]{EFEFEF}\textbf{Logging}} & \multicolumn{5}{c|}{\cellcolor[HTML]{EFEFEF}\textbf{Configuration}} & \multicolumn{4}{c}{\cellcolor[HTML]{EFEFEF}\textbf{Utility related}}
        \\ \hline
        \rowcolor[HTML]{EFEFEF} 
        \multicolumn{1}{l}{\cellcolor[HTML]{EFEFEF}\rotatebox[origin=c]{90}{\textbf{Keras}}} & \multicolumn{1}{l}{\cellcolor[HTML]{EFEFEF}\rotatebox[origin=c]{90}{\textbf{TensorFlow}}} & \multicolumn{1}{c|}{\cellcolor[HTML]{EFEFEF}\rotatebox[origin=c]{90}{\textbf{PyTorch}}}& \rotatebox[origin=c]{90}{\textbf{\% projects}} & \multicolumn{1}{l}{\cellcolor[HTML]{EFEFEF}\rotatebox[origin=c]{90}{\textbf{Keras}}} & \multicolumn{1}{l}{\cellcolor[HTML]{EFEFEF}\rotatebox[origin=c]{90}{\textbf{PyTorch}}} & \multicolumn{1}{c|}{\cellcolor[HTML]{EFEFEF}\rotatebox[origin=c]{90}{\textbf{TensorFlow}}} & \rotatebox[origin=c]{90}{\textbf{\% projects}} & \multicolumn{1}{l}{\cellcolor[HTML]{EFEFEF}\rotatebox[origin=c]{90}{\textbf{Keras}}} & \multicolumn{1}{l}{\cellcolor[HTML]{EFEFEF}\rotatebox[origin=c]{90}{\textbf{TensorFlow}}} & \multicolumn{1}{c|}{\cellcolor[HTML]{EFEFEF}\rotatebox[origin=c]{90}{\textbf{PyTorch}}} & \rotatebox[origin=c]{90}{\textbf{\% projects}} & \multicolumn{1}{l}{\cellcolor[HTML]{EFEFEF}\rotatebox[origin=c]{90}{\textbf{TensorFlow}}} & \multicolumn{1}{l}{\cellcolor[HTML]{EFEFEF}\rotatebox[origin=c]{90}{\textbf{Caffe}}} & \multicolumn{1}{c|}{\cellcolor[HTML]{EFEFEF}\rotatebox[origin=c]{90}{\textbf{PyTorch}}} & 
        \rotatebox[origin=c]{90}{\textbf{\% projects}} & \multicolumn{1}{l}{\cellcolor[HTML]{EFEFEF}\rotatebox[origin=c]{90}{\textbf{Keras}}} & 
        \multicolumn{1}{l}{\cellcolor[HTML]{EFEFEF}\rotatebox[origin=c]{90}{\textbf{TensorFlow}}} & 
        \multicolumn{1}{l}{\cellcolor[HTML]{EFEFEF}\rotatebox[origin=c]{90}{\textbf{Caffe}}} & \multicolumn{1}{c|}{\cellcolor[HTML]{EFEFEF}\rotatebox[origin=c]{90}{\textbf{PyTorch}}} & 
        \rotatebox[origin=c]{90}{\textbf{\% projects}}& \multicolumn{1}{l}{\cellcolor[HTML]{EFEFEF}\rotatebox[origin=c]{90}{\textbf{Keras}}} & 
        \multicolumn{1}{l}{\cellcolor[HTML]{EFEFEF}\rotatebox[origin=c]{90}{\textbf{TensorFlow}}} & \multicolumn{1}{c|}{\cellcolor[HTML]{EFEFEF}\rotatebox[origin=c]{90}{\textbf{PyTorch}}} & \rotatebox[origin=c]{90}{\textbf{\% projects}} \\ \midrule
        & \checkmark & \multicolumn{1}{c|}{\checkmark} & 18\% &  & \checkmark & \multicolumn{1}{c|}{\checkmark} & 12\% & \checkmark & \checkmark & \multicolumn{1}{c|}{} & 18\% &  & \checkmark & \multicolumn{1}{c|}{\checkmark} & 0.2\% & \checkmark &  &  & \multicolumn{1}{c|}{\checkmark} & 2\% &  & \checkmark & \multicolumn{1}{c|}{\checkmark} & 14\% \\ \midrule
        \checkmark & \checkmark & \multicolumn{1}{c|}{} & 12\% & \checkmark &  & \multicolumn{1}{c|}{\checkmark} & 4\% &  & \checkmark & \multicolumn{1}{c|}{\checkmark} & 13\% & \checkmark &  & \multicolumn{1}{c|}{\checkmark} & 0.2\% & \checkmark & \checkmark &  & \multicolumn{1}{c|}{} & 2\% & \checkmark & \checkmark & \multicolumn{1}{c|}{} & 14\% \\ \midrule
        \checkmark & \checkmark & \multicolumn{1}{c|}{\checkmark} & 4\% & \checkmark & \checkmark & \multicolumn{1}{c|}{\checkmark} & 1\% & \checkmark &  & \multicolumn{1}{c|}{\checkmark} & 4\% &  &  & \multicolumn{1}{c|}{} &  &  &  & - & \multicolumn{1}{c|}{\checkmark} & 0.5\% & \checkmark & \checkmark & \multicolumn{1}{c|}{\checkmark} & 7\% \\ \midrule
        \checkmark &  & \multicolumn{1}{c|}{\checkmark} & 2\% & \checkmark & \checkmark & \multicolumn{1}{c|}{} & 0.2\% & \checkmark &  & \multicolumn{1}{c|}{\checkmark} & 3\% &  &  & \multicolumn{1}{c|}{} &  &  & \checkmark &  & \multicolumn{1}{c|}{\checkmark} & 0.5\% & \checkmark &  & \multicolumn{1}{c|}{\checkmark} & 2\% \\ \midrule
        \end{tabular}
}
\label{tab:frameworkrelated}
\end{table*}

\textbf{From table~\ref{tab:modelRelated}, we highlight the most frequent combination is between \texttt{Keras} $\Leftrightarrow$ \texttt{TensorFlow} in the model training and model evaluation stages with respective percentage of project of 33\% and 6\%. Also, we illustrate that the combination between \texttt{PyTorch} $\Leftrightarrow$ \texttt{TensorFlow} is the most frequent in the model deployment and distribution computation stages}. In Table~\ref{tab:modelRelated}, we present the different DL library combinations in the model-related category. We observe that the combination \texttt{Keras} $\Leftrightarrow$ \texttt{TensorFlow} is present across the four stages in the model training category, however, it is the most frequent in the Model training stage with 33\% of the projects. This is explained by the fact that Keras is built on top of Tensorflow which makes it easy to use them together. Moreover, from Table~\ref{tab:modelRelated}, the model training stage shows a high dependency between DL libraries compared to other stages, Where the percentage of the projects with dependent DL libraries is at least three times greater than the rest of the stages (e,i. 33\% model training, 11\% model deployment). Moreover, We observe that the combination \texttt{PyTorch} $\Leftrightarrow$ \texttt{TensorFlow} is present in the first model training stage in 24\% of the projects, second in the model deployment stage with the highest appearance in 11\% of the projects, and finally in Distribution computation stage in 4\% of the projects.   

\textbf{From table~\ref{tab:frameworkrelated}, we highlight the most frequent combination is between \texttt{PyTorch} $\Leftrightarrow$ \texttt{TensorFlow} in the mathematical calculation and data type and structure sub$-$category with respective percentage of project of 18\% and 12\%. Also, we see that the combination between \texttt{Keras} $\Leftrightarrow$ \texttt{TensorFlow} is the most frequent in the tensor and array manipulation sub$-$category}. In Table~\ref{tab:frameworkrelated}, we present the different library combination in the framework related category and utility related category. In the \texttt{Framework related category}, five of the seven sub-category identified during the manual labeling process show a dependency between DL libraries. From Table~\ref{tab:frameworkrelated}, the combination \texttt{PyTorch} $\Leftrightarrow$ \texttt{TensorFlow} is omnipresent in all the five sub-categories. In fact, it is the most frequent in the sub-categories \texttt{mathematical calculation} and \texttt{data type and structure} with respective project percentage of 18\% and 12\%, while it is the second most recurrent combination in the \texttt{Tensor and array manipulation} with a project percentage of 13\%. Similarly to previous categories, we identify the dependency between \texttt{Keras} $\Leftrightarrow$ \texttt{TensorFlow} in the categories \texttt{Tensor and array manipulation}, \texttt{mathematical calculation} and \texttt{data type and structure} with respective project percentage of 18\%, 12\% and 4\%. Similar to the data-processing category, we identify a direct dependency between \texttt{PyTorch} $\Leftrightarrow$ \texttt{TensorFlow} $\Leftrightarrow$ \texttt{Keras} in the sub-categories \texttt{mathematical calculation} and \texttt{data type and structure} (see Table~\ref{tab:frameworkrelated}). 

\textbf{From table~\ref{tab:frameworkrelated}, we highlight the most frequent combination is between \texttt{PyTorch} $\Leftrightarrow$ \texttt{TensorFlow} and \texttt{Keras} $\Leftrightarrow$ \texttt{TensorFlow} with the same percentage of projects (14\%).} In Table~\ref{tab:frameworkrelated}, similar to all previous categories, the utility related category show a dependency between \texttt{PyTorch} $\Leftrightarrow$ \texttt{TensorFlow} and \texttt{PyTorch} $\Leftrightarrow$ \texttt{TensorFlow} with project percentage of 14\% in both combination. Moreover, we detect 7\% of the projects have a dependency between \texttt{PyTorch} $\Leftrightarrow$ \texttt{TensorFlow} $\Leftrightarrow$ \texttt{Keras} at the same time.

\begin{mybox}{RQ3 1/2: \RQThree}

     In particular:
    \begin{itemize}[itemsep = 3pt, label=\textbullet, wide = 0pt]
      \item Practitioners have a tendency tend to show a dependency between \texttt{Keras}$\Leftrightarrow$\texttt{Tensorflow} and \texttt{Scikit$-$learn} $\Leftrightarrow$ \texttt{PyTorch} in the \texttt{Data collection} stage, while in the \texttt{Feature engineering} stage the dependency between \texttt{PyTorch} $\Leftrightarrow$ \texttt{TensorFlow} and \texttt{Scikit$-$learn} $\Leftrightarrow$ \texttt{PyTorch} is the most frequent.
      \item \texttt{Data cleaning} stage shows a low-level of dependency with only 0.2\% of the project using multiple DL libraries.
      \item The most frequent combination is between \texttt{Keras} $\Leftrightarrow$ \texttt{TensorFlow} in the model training and model evaluation stages with respective percentage of occurrence in project of 33\% and 6\%. Also, the combination between \texttt{PyTorch} $\Leftrightarrow$ \texttt{TensorFlow} is the most frequent in the model deployment and distribution computation stages.
      \item The combination \texttt{PyTorch} $\Leftrightarrow$ \texttt{TensorFlow} is present in all the sub-categories of the framework related category.
      
    \end{itemize}
\end{mybox}

\subsubsubsection{\textbf{Dependencies combination calls a pattern of DL libraries}}
In Table~\ref{table:kerasTfRules}, we highlight the function calls association rules between the different DL library combinations across the machine learning pipeline. We report the support, as defined in the Section~\ref{subsec:assocaitionRule}, to show the importance of the rule. 

\noindent \textbf{Association rules between \texttt{Keras}$\Leftrightarrow$\texttt{TensorFlow}:} In table~\ref{table:kerasTfRules} we observe the function calls association rules between \texttt{Keras} $\Leftrightarrow$ \texttt{TensorFlow}. Firstly, in the \textbf{\texttt{model related category}}, the most relevant rule is \texttt{Keras.Sequential} $\Rightarrow$ \texttt{tf.test.main.Sequential} with support of 0.05. From the rules in the model-related category, practitioners tend to build their models using Keras while they rely on \texttt{Tensorflow} to build and test the model. Also, to compute the model in a distributed strategy, practitioners use the calls of \texttt{tf.distribute.combinations*}. For example, in the Listing~\ref{lst:exp1}(a), we see that in the repository kpe/bert-for-tf2\footnote{\url{https://github.com/kpe/bert-for-tf2}} developers  used the function \texttt{tf.execute\_eagerly()} to set up the execution of the model (e,i. eager execution) while they build the model using \texttt{Keras} (i,e. Keras.layers.Input). Second, in the \textbf{data pre-processing}, we find only one rule between \texttt{keras} and \texttt{Tensorflow} expressed as follows: \texttt{keras.Model} $\Rightarrow$ \texttt{Tensorflow.Dataset} \texttt{from\_tensor\_slices} with support of 0.013. Finally, in the \textbf{\texttt{framework-related category}}, we observe the rule \texttt{Tensorflow.python.ops.math\_ops.cast $\Rightarrow$ keras.backend.floatx} with the highest support 0.017. For example, in the listing~\ref{lst:exp1}(b) practitioner uses the function \texttt{floatX} in Keras to convert float into a string for mathematical operation with TensorFlow \texttt{math\_ops.cast} function. Besides, developers have similar behavior in testing where usually the function \texttt{testing\_utils.layer\_test} is used with \texttt{Tensorflow.test.main} to execute the unit tests because the Keras library does not offer such functionality.   
\begin{figure}[!ht]
\captionsetup{type=lstlisting}
\begin{sublstlisting}{\linewidth}

\begin{lstlisting}[language=Python]
#Repository: kpe/bert-for-tf2
#File: kpe/bert-for-tf2/blob/master/tests/test_albert_create.py
#Class: AlbertTest
def setUp(self) -> None:
    tf.compat.v1.reset_default_graph()
    tf.compat.v1.enable_eager_execution()
    +print("Eager Execution:", tf.executing_eagerly())
    
def to_model(bert_params):
    l_bert = bert.BertModelLayer.from_params(bert_params)
    +token_ids = keras.layers.Input(shape=(21,))
    ...
    return model
\end{lstlisting}
\caption{Model-related code example of the association rule \texttt{keras.layers.* $\Rightarrow$ (Tensorflow.test.main\, Tensorflow.executing\_eagerly)} as reported in code snippet line 7 and 12}
\end{sublstlisting}

\begin{sublstlisting}{\linewidth}

\begin{lstlisting}[language=Python]
#Repository: /Xilinx/Vitis-AI
#File: src/Vitis-AI-Quantizer/vai_q_tensorflow1.x/tensorflow/python/keras/layers/noise.py
#Class: AlphaDropout
def dropped_inputs(inputs=inputs, rate=self.rate, seed=self.seed): 
    ...
    kept_idx = math_ops.greater_equal(
        K.random_uniform(noise_shape, seed=seed), rate)
    +kept_idx = math_ops.cast(kept_idx, K.floatx())...
\end{lstlisting}
\caption{Framework-related code example of the association rule \texttt{Tensorflow.python.ops.math\_ops.cast $\Rightarrow$ keras.backend.floatx} as reported in code snippet line 11}
\end{sublstlisting}



\caption{Keras and Tensorflow association rules examples}
\label{lst:exp1}
\end{figure}

\noindent \textbf{Association rules between \texttt{PyTorch}$\Leftrightarrow$\texttt{TensorFlow}:} In table~\ref{table:kerasTfRules} we highlight the function calls association rules between \texttt{PyTorch} $\Leftrightarrow$  \texttt{Tensorflow}. Firstly, in the category \texttt{\textbf{model-related}}, the most relevant rule is \texttt{tensorflow.train.load\_variable $\Rightarrow$ torch.from\_numpy} with a support of 0.03. Practitioners tend to use the function \texttt{torch.from\_numpy} to convert an array into a tensor with the function load\_variable that returns a \texttt{numpy ndarray} even though \texttt{Tensorflow} provides the function \texttt{tf.convert\_to\_tensor} which converts various types to Tensor. Developers prefer to use a stray forward function such us \texttt{from\_numpy} of PyTorch instead of a general function. In the listing~\ref{lst:exp2}, we present an example of the rule where the developer an array of \texttt{Tensorflow} model variables using the \texttt{from\_numpy} PyTorch's function. Moreover, developers use the function \texttt{torch.save} with Tensorflow to save an object to a disk file because Tensorflow does not provide a function for this purpose only until recently in \texttt{Tensorflow 2.9.1}. Finally, in the category \texttt{\textbf{Framework-related}}, in \texttt{PyTorch} you can pass tensors directly to modules without a placeholder, but from the rule \texttt{torch.Tensor $\Rightarrow$ tensorflow.compat.v1.placeholder} in  table~\ref{table:kerasTfRules} developer tend to use the \texttt{tf.placeholder} function in PyTorch.    
\begin{figure}[!ht]
\captionsetup{type=lstlisting}
\begin{lstlisting}[language=Python]
#Repository: mlcommons/inference
#File: language/bert/bert_tf_to_pytorch.py
def load_from_tf(config, tf_path):..
    +init_vars = tf.train.list_variables(tf_path)
    ...
    for name, shape in init_vars:
        # print("Loading TF weight {} with shape {}".format(name, shape))
        +array = tf.train.load_variable(tf_path, name)
        ...
        print("Initialize PyTorch weight {}".format(name))
        +pointer.data = torch.from_numpy(array)
    model.qa_outputs = model.classifier
    del model.classifier
    return model
\end{lstlisting}
\caption{Model-related code example of the association rule \texttt{tensorflow.train.load\_variable $\Rightarrow$ torch.from\_numpy } as reported in code snippet line 7 and 12}
\label{lst:exp2}
\end{figure}

\begin{table*}[h]
      \caption{Example of function calls association rules between DL libraries}
    \centering
    \scalebox{0.7}{
\begin{tabular}{c
>{\columncolor[HTML]{EFEFEF}}c l|l}
\cline{3-4}
\multicolumn{1}{l}{}                                                                                           & \cellcolor[HTML]{FFFFFF}                                                                            & \multicolumn{1}{c|}{\cellcolor[HTML]{EFEFEF}\textbf{Association rule}}                                                                                                                                                                                                        & \cellcolor[HTML]{EFEFEF}\textbf{Support} \\ \midrule
\multicolumn{1}{c|}{\cellcolor[HTML]{EFEFEF}}                                                                  & \multicolumn{1}{c|}{\cellcolor[HTML]{EFEFEF}}                                                       & Keras.Sequencial $\Rightarrow$ Tensorflow.test.main                                                                                                                                                                                                                          & 0.05                                     \\
\multicolumn{1}{c|}{\cellcolor[HTML]{EFEFEF}}                                                                  & \multicolumn{1}{c|}{\cellcolor[HTML]{EFEFEF}}                                                       & keras.layers.*  $\Rightarrow$ ( Tensorflow.test.main, Tensorflow.executing\_eagerly)                                                                                                                                                                                         & 0.04                                     \\
\multicolumn{1}{c|}{\cellcolor[HTML]{EFEFEF}}                                                                  & \multicolumn{1}{c|}{\multirow{-3}{*}{\cellcolor[HTML]{EFEFEF}\textbf{Model related}}}               & Keras.Model $\Rightarrow$ tensorflow.python.distribute.combinations.*                                                                                                                                                                                                        & 0.013                                    \\ \cline{2-4} 
\multicolumn{1}{c|}{\cellcolor[HTML]{EFEFEF}}                                                                  & \multicolumn{1}{c|}{\cellcolor[HTML]{EFEFEF}}                                                       & Tensorflow.python.ops.math\_ops.cast $\Rightarrow$ keras.backend.floatx                                                                                                                                                                                                      & 0.017                                    \\
\multicolumn{1}{c|}{\cellcolor[HTML]{EFEFEF}}                                                                  & \multicolumn{1}{c|}{\multirow{-2}{*}{\cellcolor[HTML]{EFEFEF}\textbf{Data pre-processing related}}} & keras.Model $\Rightarrow$ Tensorflow.Dataset from\_tensor\_slices                                                                                                                                                                                                            & 0013                                     \\ \cline{2-4} 
\multicolumn{1}{c|}{\cellcolor[HTML]{EFEFEF}}                                                                  & \multicolumn{1}{c|}{\cellcolor[HTML]{EFEFEF}}                                                       & keras.testing\_utils.layer\_test $\Rightarrow$ Tensorflow.test.main                                                                                                                                                                                                          & 0.014                                    \\
\multicolumn{1}{c|}{\multirow{-7}{*}{\cellcolor[HTML]{EFEFEF}\rotatebox[origin=c]{90}{\parbox[c]{3cm}{\centering\textbf{Keras and Tensorflow}}}}}                   & \multicolumn{1}{c|}{\multirow{-2}{*}{\cellcolor[HTML]{EFEFEF}\textbf{Framework related}}}           & Tensorflow.test.main $\Rightarrow$ keras.testing\_utils.get\_test\_data                                                                                                                                                                                                      & 0.011                                    \\ \midrule
\multicolumn{1}{c|}{\cellcolor[HTML]{EFEFEF}}                                                                  & \multicolumn{1}{c|}{\cellcolor[HTML]{EFEFEF}}                                                       & tensorflow.train.load\_variable $\Rightarrow$ torch.from\_numpy                                                                                                                                                                                                              & 0.03                                     \\
\multicolumn{1}{c|}{\cellcolor[HTML]{EFEFEF}}                                                                  & \multicolumn{1}{c|}{\multirow{-2}{*}{\cellcolor[HTML]{EFEFEF}\textbf{Model related}}}               & tensorflow.train .list\_variables $\Rightarrow$ torch.ADD.torch.save                                                                                                                                                                                                         & 0.025                                    \\ \cline{2-4} 
\multicolumn{1}{c|}{\cellcolor[HTML]{EFEFEF}}                                                                  & \multicolumn{1}{c|}{\cellcolor[HTML]{EFEFEF}}                                                       & tensorflow.SparseTensor $\Rightarrow$ torch.sparse\_coo\_tensor                                                                                                                                                                                                              & 0.017                                    \\
\multicolumn{1}{c|}{\multirow{-4}{*}{\cellcolor[HTML]{EFEFEF}\rotatebox[origin=c]{90}{\parbox[c]{1cm}{\centering\textbf{PyTorch and Tensorflow}}}}}                 & \multicolumn{1}{c|}{\multirow{-2}{*}{\cellcolor[HTML]{EFEFEF}\textbf{Framework related}}}           & torch.Tensor $\Rightarrow$ tensorflow.compat.v1.placeholder                                                                                                                                                                                                                  & 0.017                                    \\ \midrule
\multicolumn{1}{c|}{\cellcolor[HTML]{EFEFEF}}                                                                  & \multicolumn{1}{c|}{\cellcolor[HTML]{EFEFEF}}                                                       & torch.onnx.export $\Rightarrow$ onnx2keras.onnx\_to\_keras                                                                                                                                                                                                         & 0.3                                      \\
\multicolumn{1}{c|}{\cellcolor[HTML]{EFEFEF}}                                                                  & \multicolumn{1}{c|}{\cellcolor[HTML]{EFEFEF}}                                                       & torch.onnx.export $\Rightarrow$ onnx2keras.check\_torch\_keras\_error                                                                                                                                                                                                        & 0.2                                      \\
\multicolumn{1}{c|}{\cellcolor[HTML]{EFEFEF}}                                                                  & \multicolumn{1}{c|}{\multirow{-3}{*}{\cellcolor[HTML]{EFEFEF}\textbf{Model related}}}               & autokeras.utils.get\_device $\Rightarrow$ torch.ADD.torch.nn.*                                                                                                                                                                                                               & 0.09                                     \\ \cline{2-4} 
\multicolumn{1}{c|}{\multirow{-4}{*}{\cellcolor[HTML]{EFEFEF}\rotatebox[origin=c]{90}{\parbox[c]{1.5cm}{\centering\textbf{PyTorch and Keras}}}}}                      & \multicolumn{1}{c|}{\cellcolor[HTML]{EFEFEF}\textbf{Framework related}}                             & autokeras.utils.get\_device $\Rightarrow$ (torch.load, torch.from\_numpy)                                                                                                                                                                                                    & 0.09                                     \\ \midrule
\multicolumn{1}{c|}{\cellcolor[HTML]{EFEFEF}}                                                                  & \multicolumn{1}{c|}{\cellcolor[HTML]{EFEFEF}}                                                       & \begin{tabular}[c]{@{}l@{}}(tensorflow.train, tensorflow.nn) $\Rightarrow$ (tensorflow.metrics.accuracy, \\                                                            tensorflow.metrics.accuracy\_score)\end{tabular}                                                      & 0.08                                     \\
\multicolumn{1}{c|}{\cellcolor[HTML]{EFEFEF}}                                                                  & \multicolumn{1}{c|}{\cellcolor[HTML]{EFEFEF}}                                                       & \begin{tabular}[c]{@{}l@{}}sklearn.model\_selection$\Rightarrow$ ( tensorflow.layers.*,  \\                                                tensorflow.train get\_global\_step,\\                                                tensorflow.estimator Estimator)\end{tabular} & 0.07                                     \\
\multicolumn{1}{c|}{\multirow{-6}{*}{\cellcolor[HTML]{EFEFEF}\rotatebox[origin=c]{90}{\parbox[c]{2cm}{\centering\textbf{Tensorflow or PyTorch and Scikit-learn}}}}} & \multicolumn{1}{c|}{\multirow{-6}{*}{\cellcolor[HTML]{EFEFEF}\textbf{Model related}}}               & \begin{tabular}[c]{@{}l@{}}sklearn.model\_selection $\Rightarrow$ ( trorch.nn.*, torch.optim.*, \\                                                 torch.autograd.Variable)\end{tabular}                                                                                     & 0.03                                     \\ \midrule
\end{tabular}
                }
    \label{table:kerasTfRules}
\end{table*}

\noindent \textbf{Association rules between \texttt{Keras}$\Leftrightarrow$\texttt{PyTorch}:} In Table~\ref{table:kerasTfRules}, we observe the function calls association rules between \texttt{Keras} $\Leftrightarrow$ \texttt{PyTorch}. Firstly, in the category \texttt{\textbf{model related}}, developers use PyTorch to convert the final model into a common file format using \texttt{onnx} to use the model with \texttt{Keras}, from table~\ref{table:kerasTfRules}, this behavior is supported by the first two association rule in this category with support of 0.3 and 0.2. Second, from the association rule in the \texttt{\textbf{framework related}} category, similar to \texttt{Tensorflow} practitioner, in keras projects, developers use the function \texttt{from\_numpy} in Pytorch to create Keras tensors with the AutoML tool \texttt{autokeras}.      

\noindent \textbf{Association rules between \texttt{Scikit$-$learn}$\Leftrightarrow$ (\texttt{PyTorch}, \texttt{Tensorflow}):} From table~\ref{table:kerasTfRules}, we observe that with \texttt{TensorFlow}, developers use \texttt{Scikit-learn} for \texttt{\textbf{model evaluation}} and \texttt{\textbf{model tuning}}. The first two rules in Table~\ref{table:kerasTfRules}, show that developers prefer to use the functions \texttt{sklearn.metrics.*} for the model evaluation even though Tensorflow provides similar functionalities such us \texttt{tensorflow.metrics.accuracy}\footnote{\url{https://www.tensorflow.org/api_docs/python/tf/keras/metrics}}. Moreover, PyTorch developers, use \texttt{Skicit-learn} for model tuning (e,i. sklearn.model\_selection\footnote{\url{https://scikit-learn.org/stable/modules/classes.html\#module-sklearn.model\_selection}}) since PyTorch does not offer functionalities for automatic evaluation and estimation of the model performance.   

\begin{mybox}{RQ3 2/2: \RQThree}
     In particular:
    \begin{itemize}[itemsep = 3pt, label=\textbullet, wide = 0pt]
      \item Practitioners prefer to use a function with one job such use using the function \texttt{form\_numpy} in Pytorch with Tensorflow library.
      \item Tensorflow practitioners often use functions with fewer arguments like \texttt{floatX} in Keras to convert a float to string instead the function \texttt{as\_string} in Tensorflow.
      \item Keras library shows a lack of function for testing the model which forces the users to call for Tensorflow function such us \texttt{Tensorflow.test.main} to execute all the unit tests
      \item Tensorflow and PyTorch practitioners, trust more the scikit-learn function for model evaluation and tuning. PyTorch developers use the model scikit-learn for tuning because there is no module for automatic tuning and evaluation.
    \end{itemize}
\end{mybox}
\section{Discussion}
\label{sec:discussion}
In this study, we have presented the diver dependency combination between various DL libraries across the machine learning pipeline and analyzed hidden patterns to understand the reasons behind the dependency calling attention to the most dependent DL libraries and using calls in the same function and files. This section discusses the impactful challenges and implications for researchers, library vendors, and hardware vendors.

\subsection{To Researchers} From figure~\ref{fig:DL_overTime}, results show that the new projects that use Python deep learning libraries are increasing exponentially between 2014 to 2019 this encourages the researcher to study the evolution and maintenance, understand the development practices, and analyze the DL library usage in the deep learning application. We conduct our study with the hope that researchers would use our rich information as a focal point to investigate deep learning software evolution and development practices. Moreover of the 1,484 deep learning projects on GitHub, 661 (45.5\%) projects use multiple DL libraries. This shows that DL practitioners tend to use multiple DL libraries in different machine learning workflow stages as shown in Table~\ref{tab:frameworkrelated}, ~\ref{tab:modelRelated}, and ~\ref{tab:dataProcessing}. This dependency between DL libraries comes with unique challenges for developers: (1) the evolution of the multi-library application, (2) multiple library migration, (3) DL model performances in the multi-library application, and (4) multi-library impact on hardware accelerators and so on. These challenges come with unique traits and open the door for more research topics to study. In this study, we offer an in-depth analysis of the DL library's dependencies compared to Dilhara et al.~\cite{10.1145/3453478} who studied the machine Learning libraries usage and evolution.

\subsection{To Library Vendors} Our results highlight that practitioners tend to avoid libraries with multiple tasks and too many arguments. Practitioners prefer to use straightforward functions as shown in table~\ref{table:kerasTfRules}. For example, practitioners use \texttt{from\_numpy} function of Pytorch with Tensorflow project instead \texttt{convert\_to\_tensor} in Tensorflow. Keras practitioner finds a lack of testing supports in the Keras library which pushes them to use Tensorflow. These findings can give insights to Library vendors for future releases and understand the practitioner behavior and the need to leverage more complete DL libraries. 
Practitioner uses the scikit-learn function for data processing, model evaluation, and tuning. These findings can help Library vendors to focus their efforts in a more productive way for future releases and turn their focus more on optimizing model training, monitoring, and deployment modules.
Our study helps Library vendors understand practitioners behaviors and DL library usage patterns and help optimize the future releases in more attentive way. 

\subsection{To Hardware Vendors} Our result highlights an unseen area from hardware developers that are focusing on optimizing their hardware for one specific DL library~\cite{intel}~\cite{IBM2}~\cite{Apple}. For example \texttt{Apple}~\cite{Apple} is optimizing TensorFlow, Keras, and Caffe while \texttt{Intel}~\cite{intel} is optimizing its CPUs for Tensorflow. In tables~\ref{tab:frameworkrelated}, ~\ref{tab:modelRelated}, ~\ref{tab:dataProcessing}, we highlight different combinations between deep learning libraries across different stages of the machine learning framework. These insights can be a support for hardware developers to define computation patterns in the multi-library environment and get an insight into which hardware developers should prioritize first to optimize.
\section{Related Work}
\label{sec:related-work}

\subsection{Studies of Deep Learning Systems Bugs}
Both Liu and al~\cite{10.1109/ASE51524.2021.9678891} and Zhang and al~\cite{8987482} studied bugs in deep learning applications. Liu and al~\cite{10.1109/ASE51524.2021.9678891} studied 175 bugs in deep learning applications that use Tensorflow while Zhang and al~\cite{8987482} extended the work and studied 715 Stack Overflow questions related not only to TensorFlow but PyTorch and Deeplearning4j. Both work identified software defect and development misuse problems like training anomaly and model migration and exposed implicit API usage constraints in Tensorflow. However, Zhang and al~\cite{8987482} pushed this study across different frameworks and platforms, similar to our study. Thung et al~\cite{8987482} and Sun et al~\cite{WOS:000428733800035} studied categories of bugs in the machine learning frameworks. Thung et al~\cite{8987482} scanned 500 bugs in machine learning frameworks and categorized them based on a previous taxonomy proposed by Seaman et al~\cite{10.1145/1414004.1414030} furthermore for each category, they analyzed the bug severity and the average time and effort to fix the bug. Sun et al~\cite{WOS:000428733800035} manually inspected 328 bugs in Scikit-learn, Paddle, and Caffe. They identified seven new bug categories and twelve fix patterns used to fix these bugs. 

Using multiple DL libraries could cause bugs because dependency conflicts, such bugs can be missed. Our study highlight the dependency between DL libraries and can shed light on new type of bugs and extend the taxonomy proposed by Seaman et al~\cite{10.1145/1414004.1414030} and Sun et al~\cite{WOS:000428733800035}. Also, in this study, we highlight that deep learning projects use different combinations of DL libraries. This show that the practitioners can be exposed to different dependency conflicts. Such conflicts can be the cause and explains several deep learning bugs.

\subsection{Studies of Testing Deep Learning Applications}
Testing deep learning applications is becoming a popular topic and attracting the attention of the testing community~\cite{DBLP:journals/corr/PeiCYJ17}~\cite{10.1109/ICSE.2019.00108}~\cite{DBLP:journals/corr/abs-1906-10742}~\cite{Xie2019DiffChaserDD}. Pie er al. propose DeepXplore~\cite{DBLP:journals/corr/PeiCYJ17} the first framework for white-box testing for testing real-world deep learning systems. They propose the neuron coverage metrics to measure the test effectiveness and propose a testing technique that efficiently detects behavioral inconsistencies between different deep learning models. Ma and al~\cite{8668044} propose DeepCT, a combinatorial testing technique dedicated to Depp learning systems along with combinatorial testing criteria for the test generation technique. They demonstrate that the proposed technique is more effective in capturing software defects in deep learning systems caused by adversarial examples. Zhan and al~\cite{8668044} propose DeepRoad which applies Generative Adversarial Networks (GAN) to transform driving scenes with different weather conditions (extreme weather conditions and real-world weather scenes). It was proposed to test DNN-based autonomous driving systems. Kim and al~\cite{10.1109/ICSE.2019.00108} proposes SADL (Surprise Adequacy for Deep Learning Systems) a new technique test criterion for deep learning systems which measure how far a new input data shifts from the statistical distribution of the training data.

Using multiple DL libraries can generate conflict deficit (Bugs). This is an important issue in the quality assurance process of package-based distributions. Unfortunately, the big number of configuration makes testing a difficult task and bugs from incompatibilities between libraries go undetected using normal tests. Pei et al., Ma et al, and Zhan et al did not consider the dependency of the deep library while testing the DL applications. By making practitioners aware of the dependency between DLs libraries can shed light to new approaches for testing DL systems and point to new bugs that we can refer to as conflict defects.


\subsection{Studies of Dependency Management}
Many existing work for system dependency management were proposed~\cite{7832914}. Wittern et al~\cite{7832914} studied the evolution of the npm JavaScript software ecosystems and analyse the dependency, popularity and versions distribution overtime nonetheless, we study the evolution and the popularity of the DL libraries overtime. Decan et al~\cite{DBLP:journals/corr/abs-1710-04936} studied dependency network evolution in seven software packaging ecosystems while Kikas et al~\cite{7962360} studied the structure and dependency network in three libraries ecosystem. These works highlight that network dependency is growing overtime while in this work we focus on studying the network dependency of the DL libraries. Artho et al~\cite{6224274} studied the conflict deficit of python libraries and grouped them into five main categories. Ying et al~\cite{4359473} proposed an approach to automatically detect the dependency conflicts in the python libraries, however in our study we mainly focused on studying the dependency patterns in python DL libraries.

In this work, we focus on presenting the dependency network between the deep learning libraries and how they evolve overtime. First, we highlight the evolution of the DL library usage over time. Second, we prove the dependency between 6 DL libraries on two different levels, \texttt{file level} and \texttt{function level}. Finally, we observe the distribution of different DL libraries combination over the machine learning workflow stages and highlight examples for each stage along with reasons for the dependency.
\section{Threats to Validity}
\label{sec:threats}
\subsection{Internal Validity:}
The liability of our results depends highly on the accuracy of the usage of the DL libraries in the studied projects and of extracting library calls. Due to the dynamic nature of Python, it is hard to differentiate similar constructs. To mitigate this threat, we used a state-of-the-art tool for our static analysis: Python standard AST parser for parsing the code source. To further reduce this threat, we meticulously tested the extracted data with the unit and end-to-to-end tests.
\subsection{External Validity:}
Out of the 3,143 Python deep learning projects on GitHub, we look in-depth at all the projects that use at least two deep learning libraries. The studied project accounts for a variety of domain applications, making our results can be generalized to other projects in similar domains. Nonetheless, we focused our study on Python deep learning projects, therefore, practitioners using different languages such as \texttt{C++} or \texttt{Scala} could show different behavior, challenges, and dependencies patterns when developing their solutions. However, Python has become the  most popular programming language for deep learning applications~\cite{2016introduction}~\cite{10.1145/3196398.3196445}.

\subsection{Construct validity:}
In our study of the dependency across ML workflow, we manually assigned the stages of the stage of the function call.
Our results may be subjective and depend on the judgment of the researchers who conducted the manual analysis. To mitigate this threat, two authors of the paper collectively conducted a manual analysis and reached a substantial agreement, indicating the reliability of the analysis results. To resolve disagreements, the third author joined them and each case was discussed until reaching a consensus. 
\subsection{Verifiability:}
This threat concerns the possibility to replicate this study. We have attempted to provide all the necessary details needed to replicate our study. We share our full replication package in~\cite{raed1234}.
\section{Conclusion}
\label{sec:conclusion}
DL practitioners use deep learning algorithms to build a state-of-the-art application that tries to mimic human behavior. This novel programming paradigm has become more popular than ever over the past years. The major reasons for this rapid growth are the DL libraries as they make it easy to build deep learning solutions and simplify the implementation of the deep learning algorithms. However, this rapid growth is in jeopardy without understanding the usage and the practices of the DL libraries. In this study, we perform an empirical study by mining 1,443 GitHub repositories and answering three research questions. We summarize our findings as follows: (1) Projects that use DL libraries show a rapidly increasing trend. (2) Tensorflow is the most popular DL library but start to show a decrease since 2019 (release of Tensorflow 2.0), (3) DL practitioner uses multiple deep learning libraries across ML workflow. (4) DL practitioner prefer using \texttt{Scikit-Learn} for data pre-processing and model evaluation. (5) DL practitioners tend to use simple functions with fewer arguments and with a straightforward purpose. (6) Keras library shows a lack of support for unit testing.

From our study, we hope to help deep learning library builder for future releases, deliver more optimized versions and better understand the DL practitioner need. Besides We look to give hardware builders some support to better focus their work. Finally, we look to inspire researchers to address unique challenges when advancing deep learning systems.

\bibliographystyle{ACM-Reference-Format}
\bibliography{main}

\end{document}